\documentclass[lettersize,journal]{IEEEtran}
\usepackage{amsmath,amsfonts,amssymb}
\usepackage{algorithm}
\usepackage{array}
\usepackage[caption=false,font=normalsize,labelfont=sf,textfont=sf]{subfig}
\usepackage{textcomp}
\usepackage{stfloats}
\usepackage{url}
\usepackage{verbatim}
\usepackage{graphicx}
\usepackage{epstopdf}
\usepackage{cite}
\usepackage[english]{babel}
\usepackage{amsthm}
\usepackage{array}
\usepackage[caption=false,font=normalsize,
labelfont=sf,textfont=sf]{subfig}
\usepackage{textcomp}
\usepackage{stfloats}
\usepackage{url}
\usepackage{balance}
\usepackage{multirow}
\usepackage{xcolor}
 \usepackage{caption}

\def\BibTeX{{\rm B\kern-.05em{\sc i\kern-.025em b}\kern-.08em
    T\kern-.1667em\lower.7ex\hbox{E}\kern-.125emX}}
\usepackage{balance}

\newtheorem{lemma}{Lemma}
\newtheorem{proposition}{Proposition}
\newtheorem{definition}{Definition}
\newtheorem{property}{Property}

\begin{document}

\title{Enhancing Multi-Stream Beamforming Through CQIs For 5G NR FDD Massive MIMO Communications: A Tuning-Free Scheme}
\author{Kai Li, Ying Li, Lei Cheng, \IEEEmembership{Member, IEEE}, and Zhi-Quan Luo, \IEEEmembership{Fellow, IEEE}
\thanks{The work of Zhi-Quan Luo was supported by the Guangdong Major Project of  Basic and Applied Basic Research (No.2023B0303000001), the Guangdong Provincial Key Laboratory of Big Data Computing, and the National Key Research and Development Project under grant 2022YFA1003900. The work of Lei Cheng was supported in part by the National Natural Science Foundation of China under Grant 62371418, in part by the Fundamental Research Funds for the Central Universities (226-2023-00012), and in part by the Zhejiang University Education Foundation Qizhen Scholar Foundation. (\textit{Corresponding author: Lei Cheng}.)}
\thanks{Kai Li and Zhi-Quan Luo are with The Chinese
University of Hong Kong, Shenzhen 518172, China, and also with Shenzhen Research Institute of Big Data, Shenzhen 518172,
China (e-mail: kaili4@link.cuhk.edu.cn; luozq@cuhk.edu.cn).}
\thanks{Ying Li is with the Department of Statistics and Actuarial Science, The University of Hong Kong, Hong Kong, SAR 999077, China (e-mail: lynnli98@connect.hku.hk).}
\thanks{Lei Cheng is with the College of Information Science and Electronic Engineering at Zhejiang University, Hangzhou 310058, China. He is also with Shenzhen Research Institute of Big Data, Shenzhen 518172, China (e-mail: lei\_cheng@zju.edu.cn).}
}

\markboth{Journal of \LaTeX\ Class Files,~Vol.~14, No.~8, August~2021}%
{Shell \MakeLowercase{\textit{et al.}}: A Sample Article Using IEEEtran.cls for IEEE Journals}

\IEEEpubid{0000--0000/00\$00.00~\copyright~2021 IEEE}

\maketitle

\begin{abstract}
 In the fifth-generation new radio (5G NR) frequency division duplex (FDD) massive multiple-input and multiple-output (MIMO) systems, downlink beamforming relies on the acquisition of downlink channel state information (CSI). Codebook based limited feedback schemes have been proposed and widely used in practice to recover the downlink CSI with low communication overhead. In such schemes, the performance of downlink beamforming is determined by the codebook design and the codebook indicator feedback. However, limited by the quantization quality of the codebook, directly utilizing the codeword indicated by the feedback as the beamforming vector cannot achieve high performance. Therefore, other feedback values, such as channel qualification indicator (CQI), should be considered to enhance beamforming. In this paper, we present the relation between CQI and the optimal beamforming vectors, based on which an empirical Bayes based intelligent tuning-free algorithm is devised to learn the optimal beamforming vector and the associated regularization parameter. The proposed algorithm can handle different communication scenarios of MIMO systems, including single stream and multiple streams data transmission scenarios. Numerical results have shown the excellent performance of the proposed algorithm in terms of both beamforming vector acquisition and regularization parameter learning. 
\end{abstract}


\section{Introduction}

\subsection{Background and Challenge}
\IEEEPARstart{R}{ecently}, growing demands for higher data rates, higher network capacity, and better quality of service have led to the rapid development of wireless communications, particularly in the evolution of the fifth-generation (5G) cellular systems and beyond (e.g., 6G \cite{8782879}). However, this evolution faces several challenges, such as temporally and spatially varying wireless communication environments, the limited availability of radio frequency spectrum, etc \cite{rappaport2015wideband}.  As a potential solution to these issues, massive multiple-input multiple-output (MIMO) has gained significant attention in academia and industry\cite{lu2014overview}. By deploying and harnessing a large number of antennas at the base station (BS) and the user equipment (UE), massive MIMO communication systems can efficiently mitigate the inter-cell interference \cite{MIMO_2} and utilize spectral resources \cite{MIMO_1} via some techniques. Hence, the massive MIMO communication systems can complete more complicated tasks such as channel tracking \cite{MIMO_5}, optimal user scheduling \cite{MIMO_4}, and so forth. 

Beamforming is one of the most crucial technologies in massive MIMO systems\cite{lu2014overview}. By modifying the phase and amplitude of signals during baseband processing before radio frequency transmission, this technique directs signals toward a certain direction and reduces power leakage in other directions. Therefore, beamforming can mitigate multi-user interference and gain the system capacity. In order to achieve effective beamforming, most methods assume the accurate downlink channel state information (CSI) at the BS\cite{CE1, CE2, CE3, CE4, CE5, rui}. In time division duplex (TDD) systems, the channel reciprocity enables recovering downlink CSI through the uplink channel in the BS, while frequency division duplex (FDD) systems can not achieve this recovery due to the lack of channel reciprocity \cite{wireless_book}. Hence, to obtain downlink CSI and reduce significant communication overhead from direct downlink CSI transmission, the existing 5G new radio (NR) standard proposes codebook based limited feedback schemes to encode and transmit downlink CSI in FDD systems. 

\IEEEpubidadjcol
In the context of the limited feedback scheme \cite{standard}, a shared codebook is employed by both the BS and UE. Once the UE has estimated the downlink CSI, this codebook is utilized to encode the CSI into some bits. The significance of this process lies in its ability to reduce the communication overhead, as only these bits are fed back to the BS instead of the complete original CSI. However, such information compression leads to difficulty in recovering CSI at the BS. In addition, due to the requirement of synchronous multiple data stream transmission in practice, the difficulty of recovering CSI will be further magnified. Therefore, the primary focus of this paper is to address these challenges in obtaining high-quality CSI with limited feedback bits. The recovered CSI will, in turn, be harnessed to achieve high-performance beamforming.

\subsection{Related Works}
Two kinds of techniques are actively studied for communication systems: {\it adaptive beamforming} and {\it codebook-based beamforming} \cite{6798744}.

{\it \underline{Adaptive Beamforming} :} Adaptive beamforming is a technique where the BS uses physical features of the communication system, such as the angle of arrival, to recover CSI. With this information, the BS can calculate the beam direction to maximize the beamforming gain. Although this beamforming technique is capable of maximizing the directive gain \cite{yong201160ghz, celik2008implementation}, the computational complexity and overhead of recovering CSI make it almost infeasible for popular massive MIMO systems. Due to a large number of antennas deployed in the massive MIMO system, the corresponding CSI is significantly complex, resulting in a substantial overhead of recovering CSI \cite{ramachandran2010adaptive}. The overhead also increases the beamforming setup time, which severely hinders the application of this technology in delay-sensitive communication networks. Nevertheless, the immense potential of this technique for enhancing communication quality continues to attract more attention.

{\it \underline{Codebook Based Beamforming} :} An alternative structured codebook based beamforming technique was first proposed in \cite{wang2009beam}, aiming to reduce the CSI recovering overhead of the adaptive beamforming technique. This technique utilizes a codebook composed of multiple beamforming vectors (a.k.a. codewords) that correspond to different beam patterns or directions. CSI can be effectively encoded and roughly recovered via these pre-defined codewords. Due to their simplicity, low overhead, and scalability, codebook based beamforming techniques have been widely adopted in communication standards \cite{code1}. For instance, \cite{au2007performance} generates independent codewords from a uniform distribution. However, such a simple codebook design does not meet the complicated requirements of practical communication systems. In \cite{hung2015low}, the discrete Fourier transform (DFT) matrix is taken as the orthogonal beamforming codebook to eliminate the efforts of searching for the angle of departure (AoD). Such a codebook is suitable for a two-dimensional (2D) channel environment, while the real channel is more complicated. In \cite{yuan2013separate}, a codebook is designed based on the more popular 3D MIMO channel model. The vertical dimension codebook is first designed. Then, combining the proposed vertical beamforming codebook and legacy horizontal dimension beamforming codebook, a 3D MIMO beamforming scheme is proposed. In particular, 3GPP standard \cite{code1} defines Type I codebook, which has been widely deployed in typical massive MIMO systems dating back to the 4G era. The payload of this codebook is low, only dozens of bits, resulting in the existence of numerous communication devices equipped with this codebook. Hence, it is crucial to focus on the systems with Type I codebook. 

In general, beamforming acquisition schemes rely on the codewords indicated by the indicators (a.k.a., precoder matrix indicator (PMI)) fed back from the UE \cite{au2007performance,hung2015low,zhou2012efficient,yuan2013separate}. These indicated codewords are directly used to determine the beam directions. However, since the finite codewords in the codebook can not cover infinite beam directions, the performance of these beamforming schemes is limited. To enhance codebook-based beamforming performance, we focus on exploiting other feedback information calculated via codebook, such as channel qualification indicator (CQI), to adaptively compute beamforming vectors. 

\subsection{Contributions}
In this paper, we consider a FDD massive MIMO system with Type I codebook and make the first attempt to seek the enhanced beamforming vectors by delving deeper into the Type I feedback information. The major contributions of this paper are summarized as follows.

\vspace{0.1cm}
\noindent $\bullet$ \textbf{Principled Problem Formulation.} 
In the FDD MIMO system with Type I codebook, beamforming performance relies on the accuracy of CSI hidden in limited feedback values, including PMI and CQI. This paper aims to improve the beamforming performance by exploiting the relation between CQIs and the optimal beamforming vectors, established through the downlink channel covariance matrix (DCCM). Specifically, the eigenvectors of DCCM determine the optimal beamforming vectors, and the CQI value is related to the eigenvalues of DCCM. Based on this relation, we propose a principled problem for obtaining high-performance beamforming.

\vspace{0.1cm}
\noindent $\bullet$ \textbf{Bayesian Based Tuning-Free Algorithm.}
The proposed beamforming estimation problem involves adopting a regularization scheme to adjust the parameters related to the unknown eigenvalues of DCCM. However, tuning these parameters could be time-consuming, which prompts us to develop a tuning-free algorithm within the Bayesian learning framework \cite{ml}. This algorithm is capable of learning the nearly optimal regularization parameters.  

\vspace{0.1cm}
\noindent $\bullet$ \textbf{The High Performance of Proposed Algorithm for Different Data Transmission Scenarios.}
This paper considers different data transmission scenarios, including single-stream and multi-stream scenarios. We first propose an effective algorithm for the single-stream scenario, and then extend it to a multi-stream algorithm by splitting the multi-stream case into several correlated single-stream cases. Numerical results using 5G NR Type I codebook and the channel samples from QUAsi Deterministic RadIo channel GenerAtor (QuaDRiGa)\footnote{https://quadriga-channel-model.de.} have shown the significant advantage of the proposed algorithm in terms of optimal beamforming vector acquisition. 

Moreover, in the conference version \cite{SPAWC}, we mainly introduce the algorithm for the single-stream scenario and omit all proofs for the algorithm development due to the page limit. This journal version provides all those omitted proofs and supplies the algorithm for the multi-stream scenario.

\subsection{Structure of This Paper and Notations}

The remainder of this paper is organized as follows. In Section \ref{sec:system_model_and_beamforming}, we present the system model and introduce the concept of beamforming. Section \ref{sec:relation} elaborates on the relationship between CQI and optimal beamforming vectors. The problem formulation is detailed in Section \ref{sec:formulation}, followed by the algorithm development for both single-stream and multi-stream cases. Section \ref{sec:numerical} provides numerical results and discussions, and we conclude in Section \ref{sec:conclusion}.

 Throughout the paper, we use boldface uppercase letters, boldface lowercase letters, and lowercase letters to respectively denote matrices, column vectors, and numbers. The superscript $(\cdot)^H$ is adopted to denote the Hermitian (conjugate) transpose matrix operator. The superscript $(\cdot)^\star$ is used to indicate the optimal solution. The set containing $N$ elements is denoted by $\{(\cdot)_n\}_{n=1}^N$, where $n$ is the index of the element. In the context of the complex number field, the operator $|\cdot|$ is defined as $\sqrt{\Re\{\cdot\}^2 + \Im\{\cdot\}^2}$, while in the real number field, it degenerates into the absolute value operator. The complex conjugate is represented by $\overline{(\cdot)}$, and $\|\cdot\|$ denotes the $2$-norm. In addition, $\mathrm{Tr}(\cdot)$, $\mathbb{E}[\cdot]$, and $\mathrm{diag}(\cdot)$ denote the trace, rank, expectation, and diagonalization operators respectively.

\section{System Model and Beamforming}
\label{sec:system_model_and_beamforming}
This section provides the preliminary knowledge. It splits the communication process into two phases.  Section \ref{subsec:feedback_scheme} introduces the channel state information reference signals (CSI-RS)  based limited feedback scheme, which helps the BS to acquire downlink CSI as further achieve beamforming. Then, we delve into the beamforming technique in the second phase, as discussed in Section \ref{subsec:beamforming}. 

 It is worth noting that single-user beamforming serves as the foundation for achieving multi-user beamforming in practical massive MIMO systems. In the ideal scenario with an infinite number of antennas, massive MIMO systems can achieve perfect spatial multiplexing of multiple users\cite{verenzuela1999massive}. Consequently, the multi-user beamforming scheme simply combines multiple single-user beamforming schemes. For simplicity, we mainly focus on the single-user case in this paper\footnote{Despite practical limitations on the number of antennas preventing the realization of perfect spatial multiplex, the system can mitigate interference from multiple UEs sharing the same time-frequency resource by assigning different weights to individual single-user beamforming.}. 

\subsection{Limited Feedback Scheme}
\label{subsec:feedback_scheme}
\begin{figure}[!tbp]
    \centering
    \includegraphics[width= 3.5in]{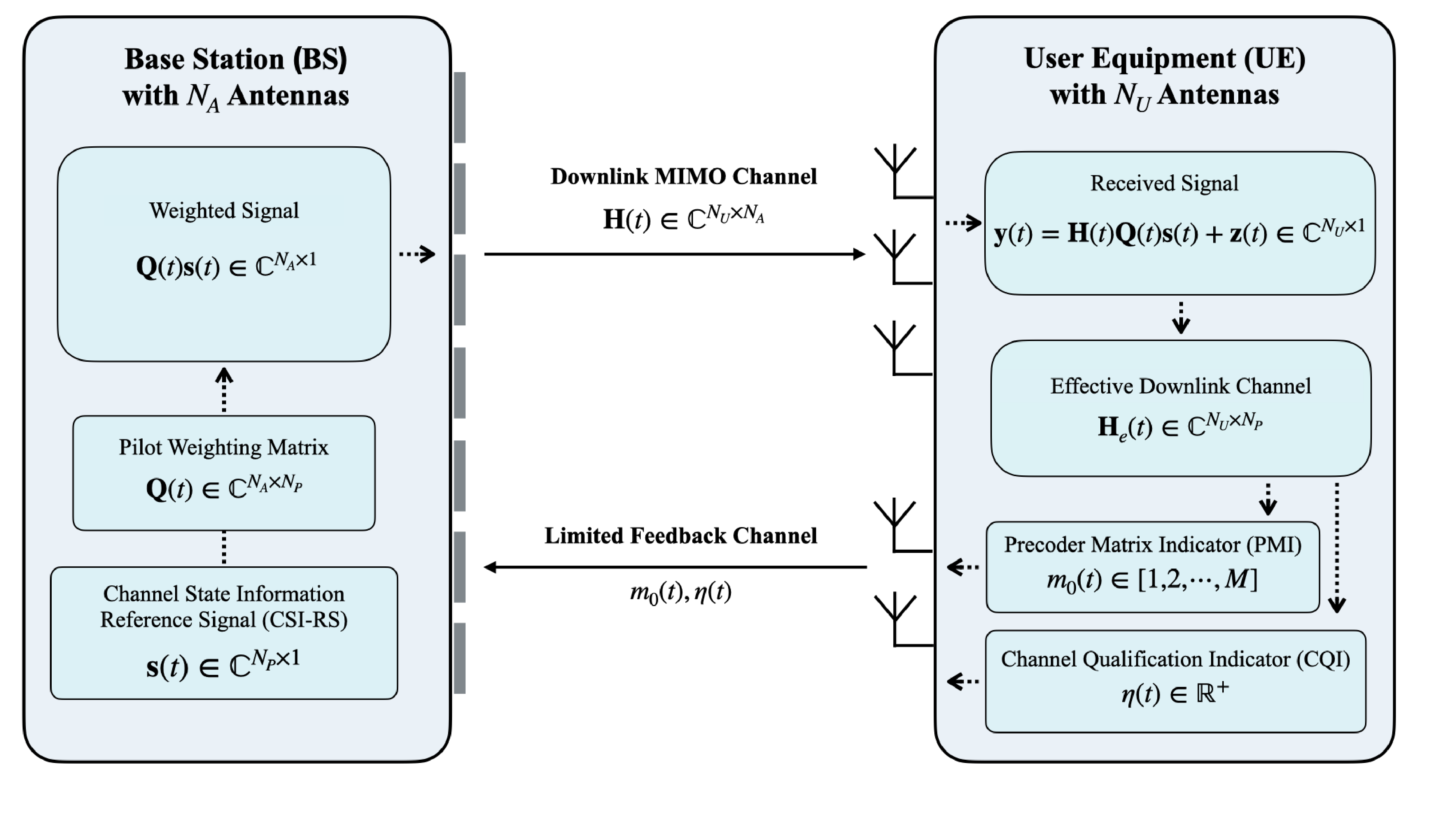}
 
    \caption{Block diagram of downlink transmission and feedback mechanism.}
    \label{fig:system_model}
\end{figure}
Consider a point-to-point MIMO link consisting of a BS with $N_A$ antennas and a UE with $N_U$ antennas, as shown in Fig.\ref{fig:system_model}. At the communication round $t$, the downlink MIMO channel matrix between the BS and the UE is denoted by $\mathbf H(t) \in \mathbb C^{N_U \times N_A}$ ($N_U\leq N_A$) with the corresponding downlink covariance matrix $ \mathbf{C}(t) = \mathbf H(t)^H \mathbf H(t) \in \mathbb C^{N_A \times N_A}$. Within $T$ consecutive communication rounds, we assume the channel remains stationary, \textcolor{black}{resulting in DCCM $\mathbf{C} = \mathbb{E}_t(\mathbf{C}(t))$ remaining unchanged.}

5G NR standard \cite{code1} adopts the special channel state information reference signals (CSI-RSs) $\{\mathbf s(t) \in \mathbb C^{N_P}\}_t$ to assist acquisition of the downlink CSI at the BS, which is stated as follows.  The dimension of CSI-RS is determined by the number of ports (also known as virtual antennas), denoted by $N_P$. In general, $N_P$ is smaller than $N_A$, but more than $N_U$. At the $t$-th communication round, the BS leverages the pilot weighting matrix $\mathbf Q(t) \in \mathbb C^{N_A \times N_P}$, which is usually assumed to be semi-unitary (i.e., $\mathbf{Q}(t)^H\mathbf{Q}(t) = \mathbf{I}$), to match the dimension of the BS antennas when transmitting $\mathbf s(t)$. After transmitting the reshaped signal $\mathbf Q(t) \mathbf s(t) \in \mathbb C^{N_A}$ through the downlink channel, the UE receives a signal modeled as 
\begin{align}
    \mathbf y(t) = \mathbf H(t) \mathbf Q(t) \mathbf s(t) + \mathbf z(t), \label{1}
\end{align}
where $\mathbf z(t) \sim \mathcal {CN}(\mathbf 0, \sigma^2 \mathbf I_{N_U})$ is the additive white Gaussian noise (AWGN).   By utilizing $\mathbf s(t)$, the UE can acquire the effective downlink channel\footnote{{Due to the unknown pilot matrix $\mathbf{Q}(t)$, it is easier for the user to estimate the effective channel $ \mathbf{H}(t)\mathbf{Q}(t)$ rather than the instantaneous channel $\mathbf{H}$(t). }}
\begin{align}
    \mathbf H_e(t) = \mathbf H(t) \mathbf Q(t),  \label{3}
\end{align}
via various channel estimation algorithms \cite{channel_est_UE_1,channel_est_UE_2,channel_est_UE_3}.  

In FDD systems, the UE does not directly transmit $\mathbf H_e(t)$ back to the BS via the reverse link due to the huge overhead of communication. Instead, $\mathbf H_e(t)$ is ``encoded'' into several bits by using Type I codebook introduced by 3GPP \cite{code1}. According to the number of data streams $N_D$ ($1\leq N_D\leq N_U$), the system will adopt a corresponding $N_D$-layer Type I codebook, denoted as $\mathcal V = \{\mathbf V_m \in \mathbb C^{N_P\times N_D};\mathbf{V}^H_m\mathbf{V}_m = \mathbf{I} ~|~ m = 1,\cdots, M^{\tiny (N_D)}\}$.
By exploiting this codebook, the UE computes the PMI 
\begin{align}
  m_0(t)   &=\mathop{\arg\max}_{m=1,\cdots, M^{\tiny (N_D)}} \mathrm{Tr}\left( \mathbf V_m^H  \mathbf{H}_e(t)^H \mathbf{H}_e(t) \mathbf V_m \right), \nonumber\\
  &=\mathop{\arg\max}_{m=1,\cdots, M^{\tiny (N_D)}} \mathrm{Tr}\left( \mathbf V_m^H  \mathbf Q(t)^H \mathbf C \mathbf Q(t)  \mathbf V_m \right), \label{multi_PMI}
  \end{align}
and the CQI
\begin{align}
\eta (t) &=\mathrm{Tr} \left (\mathbf V_{m_0(t)}^H \mathbf{H}_e(t)^H \mathbf{H}_e(t)  \mathbf V_{m_0(t)}\right ) \nonumber\\
&=\mathrm{Tr} \left (\mathbf V_{m_0(t)}^H \mathbf Q(t)^H \mathbf C \mathbf Q(t)  \mathbf V_{m_0(t)}\right ).\label{multi_CQI}
\end{align} Particularly, in single-stream case (where $N_D = 1$), codeword $\mathbf{V}_m$ is simplified as vector $\mathbf{v}_m$, and correspondingly the PMI and CQI are reduced to
\begin{align}
    m_0(t) &= \mathop{\arg\max}_{m=1,\cdots,M} ~\mathbf v^H_m \mathbf Q(t)^H \mathbf C \mathbf Q(t)  \mathbf v_m, \label{PMI}\\
      \eta (t) &=\mathbf v^H_{m_0(t)} \mathbf Q(t)^H \mathbf C \mathbf Q(t) \mathbf v_{m_0(t)}\label{CQI}.
\end{align}
Subsequently, PMI and CQI are fed back to the BS with reduced overhead. This CSI transmission scheme is called the limited feedback scheme.

\subsection{Beamforming}
\label{subsec:beamforming}
 In the typical single-user multi-stream beamforming massive MIMO model, the BS transmits signals $\{x_i \in \mathbb{C}\}_{i=1}^{N_D}$ toward the UE along $N_D$ beam directions (each corresponding to an individual data stream), which are mathematically represented by orthogonal unit-norm beamforming vectors $\{\mathbf{w}_i\}_{i=1}^{N_D}$\cite{clerckx2013mimo}. Hence, at communication round $t$, the received signal at the UE is formulated as \textcolor{black}{
\begin{align}
    \mathbf{y}(t) = \mathbf{H}(t)\mathbf{W}\mathbf{x}(t) + \mathbf{z}(t), 
\end{align}
where \begin{align}
    \mathbf{W} &= [\mathbf{w}_1,\mathbf{w}_2,\cdots,\mathbf{w}_{N_D}];\\
    \mathbf{x}(t) &= \left[x_1(t),x_2(t),\cdots,x_{ N_D}(t)\right],
\end{align}
and $\mathbf z(t) \sim \mathcal {CN}(\mathbf 0, \sigma^2 \mathbf I_{N_U})$ is the AWGN.}

The quality of this transmission process is effectively measured by the average received signal-to-noise ratio (SNR), since it determines the maximum achievable data rate\footnote{In single-user multi-stream cases, all streams are transmitted towards the same user, resulting in no interference among them.}. \textcolor{black}{With the assumption that the transmitted signals $\{x_i(t)\}_{i=1}^{N_D}$ have unit power (without loss of generality), the average received SNR is given by:
 \begin{align}
    &\mathbb{E}_t\left[\frac{\sum_{i=1}^{N_D} x_{i}(t)^H \mathbf{w}_i^H\mathbf{H}(t)^H\mathbf{H}(t)\mathbf{w}_i x_i(t)}{\sigma^2}\right]\nonumber\\
    = &\mathbb{E}_t\left[\frac{\sum_{i=1}^{N_D} \left(\mathbf{w}_i^H\mathbf{H}(t)^H\mathbf{H}(t)\mathbf{w}_i\right) x_{i}(t)^H x_i(t)}{\sigma^2}\right] \nonumber\\
    = &\mathbb{E}_t\left[\mathrm{Tr}\left(\frac{\mathbf{W}^{H}\mathbf{C}(t)
   \mathbf{W}}{\sigma^{2}}\right)\right]\nonumber\\
    = &\mathrm{Tr}\left({\mathbf{W}^{H}{\bf C}
   \mathbf{W}\over\sigma^{2}}\right),\label{snr}
    \end{align}
 }where $\mathbf{C}$ is the ground-truth DCCM,  $\sigma^{2}$ is the variance of the AWGN, and an expectation over time is adopted to alleviate the effects of time-varying channels\cite{nikos}. Particularly, the equation holds in \eqref{snr} due to the stationary assumption of the channel.

From \eqref{snr}, the SNR value is determined by the beamforming vectors $\mathbf{W}$. 
\textcolor{black}{One crucial aim of beamforming is to increase the average received SNR by optimizing beamforming vectors\cite{beam_def}. This paper primarily focuses on achieving this objective. Particularly, the optimal beamforming vectors are defined as follows\cite{nikos}.
\begin{definition}
Given a point-to-point FDD MIMO link, the optimal beamforming vectors $\{\mathbf{w}^\star_i\}_{i=1}^{N_D}$ maximize the average received SNR \eqref{snr} with power budget constraint, given as
\begin{align}
        \mathbf{W}^\star =\arg\max_{\mathbf{W}^H \mathbf{W}= \mathbf{I}} \mathrm{Tr}(\mathbf{W}^H\mathbf{C}\mathbf{W}),
\end{align} 
where $\mathbf{W}^\star = [\mathbf{w}^\star_1 ~~ \mathbf{w}^\star_2 \cdots \mathbf{w}^\star_{N_U}]$. 
 \label{definition:1}
\end{definition}
Specifically, {\bf Definition \ref{definition:1}} indicates that $\{\mathbf{w}^\star_i\}_{i=1}^{N_D}$ corresponds to the first $N_D$ principal eigenvectors of the DCCM $\mathbf C$.} This dependence on the DCCM arises because instantaneous channel information is unavailable to the BS. Instead, the BS can only obtain downlink channel covariance information from PMI and CQI (see \eqref{multi_PMI} and \eqref{multi_CQI}). Nevertheless, because the ground truth $\mathbf{C}$ is unknown, the BS cannot directly compute the optimal beamforming vectors through eigen-decomposition. Instead, the practical wireless communication scheme directly uses weighted Type I codeword $\mathbf{Q}\mathbf{V}_{m_{0}}$ as the beamforming vectors. However, to minimize feedback payload, the size of Type I codebook is intentionally restricted, resulting in non-negligible quantization errors. These errors, stemming from the discrete representation of beamforming directions, significantly degrade the performance of practical beamforming methods. Consequently, it is crucial to investigate novel approaches that can leverage feedback information to enhance beamforming performance in the presence of quantization limitations.

In previous research \cite{li2022downlink}, we have proposed to estimate $\mathbf{C}$ from CQIs and PMIs. However, this approach demands a substantial amount of feedback to mitigate estimation errors. While it improves the quality of beamforming vectors, a lot of feedback requirements create an unaffordable overhead in real-world scenarios. Nevertheless, the low-rank feature of $\mathbf{C}$ implies that there is no need to estimate the entire matrix. Instead, we can directly estimate the optimal beamforming vectors by establishing a relationship between them and the feedback values, as elaborated in the following section.

\section{The Relation Between CQI and the Optimal Beamforming Vectors}
\label{sec:relation}
Considering the differentiation between single-stream and multi-stream feedback schemes, this section sequentially discusses the relationship between CQI and the optimal beamforming vectors in these two scenarios. Further details on these relationships are elaborated in Section \ref{sec:link} for the single-stream case and Section \ref{sec:multi-stream link} for the multi-stream case.

\subsection{Single-stream Scenario}
\label{sec:link}
As pointed out in \textbf{Definition \ref{definition:1}}, in the single-stream scenario, the unit-norm optimal beamforming vector $\mathbf{w}^\star$ is the first principal eigenvector of the ground-truth DCCM $\mathbf C$. In other words, $\mathbf{w}^\star$ is embedded in the eigen-decomposition of $\mathbf C$ with rank $N_U$ as follows:
\begin{align}
    \mathbf C = \sigma_1 \mathbf{w}^\star {\mathbf{w}^\star}^H + \sigma_{2} \mathbf u_{2}\mathbf u_{2}^H + \cdots + \sigma_{N_U} \mathbf u_{N_U }\mathbf u_{N_U}^H.\label{c_decomp}
\end{align}
Here, $ \left \{\mathbf{w}^\star, \{\mathbf u_k\}_{k=2}^{N_U}\right\}$ , $\{\sigma_k \}_{k=1}^{N_U}$ ($\sigma_1 \geq \sigma_2 \geq \cdots \geq \sigma_{N_U}$) are the eigenvectors and eigenvalues of $\mathbf{C}$, respectively. By substituting $\mathbf{C}$ in  \eqref{CQI} with its eigen-decomposition \eqref{c_decomp}, the relation between $\mathbf{w}^\star$ and CQI $\eta(t)$ is indicated by 
\begin{align}
     \eta (t) &=  \mathbf v_{m_0(t)}^H \mathbf Q(t)^H \left(\sigma_1 \mathbf{w}^\star {\mathbf{w}^\star}^H\right) \mathbf Q(t)  \mathbf v_{m_0(t)} + \cdots   \nonumber\\
     & ~~~~~~~~+\mathbf v_{m_0(t)}^H \mathbf Q(t)^H \left(\sigma_{N_U} \mathbf u_{N_U} \mathbf u_{N_U} ^H\right) \mathbf Q(t)  \mathbf v_{m_0(t)} \nonumber \\
     &= |\sqrt{\sigma_1} {\mathbf v_{m_0(t)}^H \mathbf Q(t)^H \mathbf{w}^\star} |^2 
     +  \sum_{k=2}^{N_U} |\sqrt{\sigma_k} \mathbf v_{m_0(t)}^H \mathbf Q(t)^H  \mathbf u_k |^2, \label{link}
\end{align}
based on which we can further conclude the following property. 
\begin{property}
     At the communication round $t$, there exists a positive constant $\gamma \geq 1$ such that
    \begin{align}
    \eta (t) = \mathbf v_{m_0(t)}^H \mathbf Q(t)^H (\gamma {\sigma}_1 \mathbf{w}^\star {\mathbf{w}^\star}^H) \mathbf Q(t)  \mathbf v_{m_0(t)}.
    \end{align} \label{prop:2}
\end{property}
\vspace{-0.5cm}
Notice that the approximation error between $\eta(t)$ and $\mathbf v_{m_0(t)}^H \mathbf Q(t)^H ( {\sigma}_1 \mathbf{w}^\star {\mathbf{w}^\star}^H) \mathbf Q(t)  \mathbf v_{m_0(t)}$ is proportional to $\gamma$, which value is influenced by several factors, including the eigenvalues of matrix $\mathbf{C}$ and the choice of $\mathbf{Q}(t)$. For example, when the eigenvalues of $\mathbf{C}$ are similar, $\gamma$ tends to be relatively large. In addition, if $\mathbf{u}_k$ lies within column space of $\mathbf{Q}(t)$ (denoted by $\mathrm{Col}(\mathbf{Q}(t))$) while $\mathbf{w}^\star$ does not, $\gamma$ also tends to be infinity. These issues would introduce a notable approximation error, resulting in $\mathbf{w}^\star$ not being well recovered based on the relation indicated by {\bf Property \ref{prop:2}}. 
 
It is worth noting that research studies, such as those presented in \cite{abouda2006effect} and \cite{abouda2005impact}, demonstrate significant variations in the eigenvalues of $\mathbf{C}$ in real-world scenarios. Hence, to avoid situations where $\gamma$ is substantially large, we consider tuning $\mathbf{Q}(t)$ such that  $\mathbf{w}^\star\in \mathrm{Col}(\mathbf{Q}(t))$. The effectiveness of such a tuning is demonstrated by the following proposition.  
\begin{proposition}
    Assuming that $\{\mathbf{w}^\star, \{\mathbf u_k\}_{k=2}^{N_U}\}$ are orthogonal random variables on sphere $\mathbb{S}^{N_A - 1}$. At the communication round $t$, if $\mathbf{w}^\star\in \mathrm{Col}(\mathbf{Q}(t))$ holds, there is
    \begin{align}
    \eta (t)  = | \sqrt{\sigma_1} \mathbf v_{m_0(t)}^H \mathbf Q(t)^H \mathbf{w}^\star|^2 + \delta_t, \label{8}
    \end{align}
where 
\begin{align}
    \delta_t =  { \sum_{k=2}^{N_U} \sigma_k |\mathbf v_{m_0(t)}^H \mathbf Q(t)^H  \mathbf u_k |^2}.
\end{align}
Regarding $\delta_t$, there exist positive constants $\epsilon_{(\text{ac})}$ and $\epsilon_{(\text{bm})}$ such that
\begin{align}
     &\mathbb{P}\left( |\delta_t  - \sum_{k=2}^{N_U}\sigma_k(1-\mu_{(\text{ac})})\mu_{(\text{bm})}| <\sum_{k=2}^{N_U}\sigma_k\epsilon \right) \nonumber\\
     &~~~~~~~~~~~~~~~~~~~~~~\geq \left((1-\frac{\sigma_{(\text{ac})}^2}{\epsilon^2_{(\text{ac})}})(1-\frac{\sigma^2_{(\text{bm})}}{\epsilon^2_{(\text{bm})}})\right)^{N_U-1},
\end{align}
\textcolor{black}{Here, $\mu_{(\text{bm})} = \frac{1}{N_A},\sigma^2_{(\text{bm})} = \frac{2(N_A-1)}{N_A^3 + 2N_A} $ are the mean and variance, respectively, of the variable $|\mathbf{o}\mathbf{Q}(t)^H\mathbf{u}_k|^2$, which follows beta distribution $\mathrm{Beta}(\frac{1}{2},\frac{N_A-1}{2})$. $\mu_{(\text{ac})}$ and  $\sigma_{(\text{ac})}^2$ are the mean and variance of variable $|\mathbf{v}_{m_0(t)}^H\mathbf{Q}(t)^H\mathbf{w}^\star|^2$, with cumulative distribution function $\left[I\left(\frac{1}{2}, \frac{N_P}{2}\right)\right]^{M}$ ($I $ is regularized incomplete beta function). Particularly, $\mathbf{o}$ is orthogonal to $\mathbf{Q}(t)^H\mathbf{w}^\star$ and their linear combination forms $\mathbf{v}_{m_0(t)}$. The term $\epsilon$ is given by $(1-\mu_{(\text{ac})})\epsilon_{(\text{bm})} + \mu_{(\text{bm})}\epsilon_{(\text{ac})} + \epsilon_{(\text{ac})}\epsilon_{(\text{bm})}$. }
    \proof see Appendix \ref{proof:proposition1}.
    \label{prop:1}
\end{proposition}
 \noindent \textcolor{black}{{\bf Proposition \ref{prop:1}} characterizes the probability distribution of the difference between matching error $\delta_t$ and value $\sum_{k=2}^{N_U}\sigma_k(1-\mu_{(\text{ac})})\mu_{(\text{bm})}$. In typical massive MIMO systems, the value  $\sum_{k=2}^{N_U}\sigma_k(1-\mu_{(\text{ac})})\mu_{(\text{bm})}$ approximates $0$, and the difference tends to $0$ with a high probability. This implies that $\delta_t$ tends to be $0$, causing $\gamma$ to approach $1$\footnote{\textcolor{black}{Consider a situation where $N_A = 128, N_P =8, N_U = 2$ and $M = 256$. Through numerical simulation, $\mu_{(\text{ac})}$ and $\sigma^2_{(\text{ac})}$ can be approximated to be $0.7447$ and $0.0066$ respectively.Additionally, we find $\mu_{(\text{bm})}=0.0078125$ and $\sigma^2_{(\text{bm})}\approx 0.000122$. It is straightforward to verify that $\sum_{k=2}^{N_U}\sigma_k(1-\mu_{(\text{ac})})\mu_{(\text{bm})} \approx 2e-3$. By selecting $\epsilon_{(\text{ac})} =0.2$ and $\epsilon_{(\text{bm})} =0.01$, the matching error $\delta_t$ approaches $2e-3$ with a precision of $4e-3$ with a probability of approximately $0.748$.}}.}

Though {\bf Proposition \ref{prop:1}} exhibits the possibility of $\gamma \rightarrow 1$, achieving the condition $\mathbf{w}^\star\in \mathrm{Col}(\mathbf{Q}(t))$ is still challenging. This difficulty arises because the base station requires more channel information to construct an appropriate weighting matrix $\mathbf{Q}(t)$, which may not be readily available in practice. Particularly, in the general case where $\mathbf{w}^\star \notin \mathrm{Col}(\mathbf{Q}(t))$, the impact of $\gamma$ cannot be overlooked. However, this influence still can be mitigated through a randomized pilot matrix generation scheme based error averaging mechanism. Specifically, this mechanism involves using multiple feedback values derived from random pilot matrices to offset the impact of $\gamma$. Building upon this idea and {\bf Property \ref{prop:2}}, an estimate scheme for $\mathbf{w}^\star$ is elaborated in Section \ref{sec:formulation_1}.

\subsection{Multi-stream Scenario}
\label{sec:multi-stream link}
In the context of the multi-stream feedback scheme, the DCCM $\mathbf C$ still links the CQI and the optimal beamforming vectors $\{\mathbf{w}^\star_i\}_{i=1}^{N_D}$. Without loss of generality, let us assume that the number of data streams $N_D$ is equal to the number of user antennas $N_U$. Consider the eigen-decomposition
\begin{align}
    \!\!\!\!\mathbf C = \sigma_1 \mathbf{w}^\star_1 {\mathbf{w}^\star_1}^H  + \sigma_{2} \mathbf w^\star_{2}{\mathbf w^\star_{2}}^H + \cdots + \sigma_{N_U} \mathbf w^\star_{N_U}{\mathbf{w}^\star_{N_U}}^{\!\!\!\!\!H},\label{multi_decoms_C}
\end{align}
where  $\{\sigma_k \}_{k=1}^{N_U}$ are the eigenvalues of $\mathbf{C}$. By substituting \eqref{multi_decoms_C} into \eqref{multi_CQI}, the relation between multi-stream CQI and the optimal beamforming vectors is indicated by
\begin{align} 
\eta (t) & = \mathrm{Trace}\left( \mathbf V_{m_0(t)}^H \mathbf Q(t)^H \left(\sum_{k=1}^{N_U}\sigma_k\mathbf{w}^\star_k {\mathbf{w}^\star_k}^H\right) \mathbf Q(t)  \mathbf V_{m_0(t)}\right) \nonumber\\
    &= \sum_{k=1}^{N_U} \underbrace{\sum_{i=1}^{N_U} \mathbf v_{i,m_0(t)}^H \mathbf Q(t)^H (\sigma_k \mathbf w^\star_k{\mathbf w^\star_k}^H) \mathbf Q(t)  \mathbf v_{i,m_0(t)}}_{\triangleq \eta_k(t)}. \label{multi_CQI_expansion}
\end{align}
which can be further decomposed into $N_U$ components, i.e., 
\begin{align}
    \eta_k(t) &= \mathbf v_{k,m_0(t)}^H \mathbf Q(t)^H (\sigma_k \mathbf w^\star_k{\mathbf w^\star_k}^H) \mathbf Q(t)  \mathbf v_{k,m_0(t)} \nonumber\\
    &~~~~~~+ \sum_{i\neq k}^{N_U} \mathbf v_{i,m_0(t)}^H \mathbf Q(t)^H (\sigma_k \mathbf w^\star_k{\mathbf w^\star_k}^H) \mathbf Q(t)  \mathbf v_{i,m_0(t)},\nonumber\\
    &~~~~~~~~~~~~~~~~~~~~~~~~~~~~~~~~~~~\forall k = 1,2,\cdots,N_U.\label{multi_CQI_expansion_component}
\end{align}

Estimating the individual values of $\{\eta_k(t)\}_{k=1}^{N_U}$ is challenging since only their summation, $\eta(t)$, is observable at the BS. Moreover, given the uncertain order of $\{\sigma_k\}_{k=1}^{N_U}$ , distinguishing the order of $\{\eta_k (t)\}_{k=1}^{N_U}$  is difficult as well. In the absence of more detailed information, we resort to a simple heuristic approach for setting the values of $\{\eta_k (t)\}_{k=1}^{N_U}$ as follows.

\noindent\emph{\underline{Average Partition of $\eta (t)$:}} $\eta(t)$ is divided equally such that 
\begin{align}
  &\eta_{k} (t) = \frac{\eta (t)}{N_U}, \quad \forall k = 1,\cdots,N_U.\label{average_par} 
\end{align}

Once $\{\eta_k(t)\}_{k=1}^{N_U}$ has been determined, based on \textbf{Proposition \ref{prop:1}} and \textbf{Property \ref{prop:2}}, we can further specify the components in \eqref{multi_CQI_expansion_component} as follows:
\begin{align}
    \eta_{k} (t) &= \mathbf v_{k,m_0(t)}^H \mathbf Q(t)^H (\hat{\sigma}_k \mathbf w^\star_k{\mathbf w^\star_k}^H) \mathbf Q(t)  \mathbf v_{k,m_0(t)},\label{link_multi}
\end{align}
where $\hat{\sigma}_k = \gamma \sigma_k$ (see $\gamma$ analysis in Section \ref{sec:link}). It is worth noticing that compared to the approximation errors in the single-stream case, the errors in \eqref{link_multi} tend to be larger, mainly due to the imprecision of the CQI partition scheme. Furthermore, the limited feedback scheme offers only a finite amount of information, making it challenging to effectively mitigate these errors.

\section{Problem Formulation and Algorithm Development}
\label{sec:formulation}
In this section, we leverage the insights presented in Section \ref{sec:relation} to model beamforming estimation problems in both single-stream and multi-stream scenarios. Subsequently, we develop algorithms to address these problems. 

To be more specific, Section \ref{sec:formulation_1} introduces the single-stream beamforming estimation problem, which we tackle using the algorithm detailed in Section \ref{sec:formulation_2}. Furthermore, in Section \ref{sec:bayes_regularization}, we outline an empirical Bayes-based mechanism for setting the regularization parameter in the single-stream beamforming estimation problem. Finally, Section \ref{sec:multi-stream} delves into the multi-stream beamforming estimation problem and extends the proposed single-stream algorithm to solve it.

\subsection{Problem Formulation}
\label{sec:formulation_1}
 
\textbf{Proposition \ref{prop:1}} and \textbf{Property \ref{prop:2}} establish a connection between CQIs and the optimal beamforming vector, which allows for the estimation of the optimal beamforming vector. Additionally, by collecting and exploiting CQIs $\{ \eta(t) \}_{t=1}^T$ and PMIs $\{m_0(t) \}_{t=1}^T$, the influence of the approximation errors outlined in Property \ref{prop:2} could be further reduced. Therefore, at the $T$-th communication round, we propose to solve the following optimization problem: 
\begin{align}
    \min_{\|\mathbf{x}\| = 1} \sum_{t=1}^{T} \left(|\sqrt{\hat \sigma_1}\mathbf{v}_{m_0(t)}^{H} \mathbf{Q}(t)^H\mathbf{x}| - \sqrt{ \eta(t)} \right)^2, \label{9}
\end{align}
where constraint $\|\mathbf{x}\| = 1$ is adopted due to the unit-norm property of $\mathbf{w}^\star$.

 However, the unknown value of $\hat \sigma_1$ in problem \eqref{9} brings an intractable difficulty. To tackle this issue, one intuitive scheme treats $\hat \sigma_1$ as another optimization variable (denoted by $\sigma$). If this scheme is adopted, the optimal solution pair $\{\sigma^\star,\mathbf{x}^\star\}$ should satisfy
\begin{align}
    |\sqrt{\sigma^\star}\mathbf{v}_{m_0(t)}^H \mathbf{Q}(t)^H\mathbf{x}^\star| = \sqrt{\eta(t)}.\label{cd1}
\end{align}
Hence, $\sigma^\star$ can be set as
$$\frac{\eta(t)}{|\mathbf{v}_{m_0(t)}^H \mathbf{Q}(t)^H\mathbf{x}^\star|^2}$$
for any given $\mathbf{x}^\star$ to make \eqref{cd1} hold, even when $\mathbf{x}^\star\neq\mathbf{w}^\star$. In other words, variable $\sigma$ overfits the noise between $|\sqrt{\hat{\sigma}_1}\mathbf{v}_{m_0(t)}^H \mathbf{Q}(t)^H\mathbf{w}^\star|$ and $\sqrt{\eta(t)}$, which causes deviation from the ground truth and further degrades the estimation of the optimal beamforming vector. Given the shortcomings of the above scheme, another scheme is adopted to remove the unknown $\hat \sigma_1$ from problem \eqref{9} via regularization. To be specific, by using variable substitution, i.e., $\mathbf{w}\triangleq \sqrt{\hat \sigma_1}\mathbf{x}$, and then relaxing equality constraint  $\|\mathbf w\| = \sqrt{\hat \sigma_1}$, problem \eqref{9} can be transformed into: 
\begin{align}
    \min_{\|\mathbf{w}\|\leq \sqrt{\hat \sigma_1}} \sum_{t=1}^{T} \left(|\mathbf{v}_{m_0(t)}^{H} \mathbf{Q}(t)^H\mathbf{w} | - \sqrt{ \eta(t)} \right)^2. \label{11}
\end{align}
Based on duality theory \cite{opt}, the constrained problem \eqref{11} can be equivalently reformulated as the following unconstrained optimization problem 
\begin{align}
  \min_{\mathbf{w}} \sum_{t=1}^{T} \left(|\mathbf{v}_{m_0(t)}^{H} \mathbf{Q}(t)^H\mathbf{w} | - \sqrt{ \eta(t)} \right)^2 + \lambda\|\mathbf{w}\|^2. \label{12}
\end{align}
 Here, the value of the regularization parameter $\lambda>0$ is related to unknown $\hat \sigma_1$, and thus resulting in the difficulty of parameter configuration (see Section \ref{sec:bayes_regularization} for more discussion). 

 In addition, the non-convexity property of modulus terms $\{|\mathbf{v}_{m_0(t)}^{H} \mathbf{Q}(t)^H\mathbf{w} |\}^T_{t=1}$ hinders solving problem \eqref{12}. Meanwhile, notice that the relation \eqref{link} remains unchanged even with the introduction of phase rotation to the term $\mathbf{v}_{m_0(t)}^{H} \mathbf{Q}(t)^H {\mathbf{w}^\star}$ due to the quadratic formulation. Inspired by these observations, we introduce auxiliary phase rotation variables $\{{\phi_t}\}^{T}_{t=1}$ and equivalently reformulate problem \eqref{12} as follows: 
\begin{align}
    \min_{\mathbf{w},{\{\phi_t\}}^{T}_{t=1}} \sum_{t=1}^{T} &\mathbf{{w}}^H \mathbf{Q}(t) \mathbf{v}_{m_0(t)}\mathbf{v}_{m_0(t)}^{H} \mathbf{Q}(t)^H {\mathbf{w}} \nonumber \\
 &- 2\sqrt{\eta(t)}  \Re\{\mathbf{v}_{m_0(t)}^{H} \mathbf{Q}(t)^H {\mathbf{w}}\exp(j\phi_t)\} + \lambda\|{\mathbf{w}}\|^2. \label{EQ:14}
\end{align}
The equivalence between problem formulations \eqref{12} and \eqref{EQ:14} is shown in Appendix \ref{proof:equivalence}. 

Since problem \eqref{EQ:14} involves decoupled optimization variables $\{\mathbf{w},\{\phi_t \}_{t=1}^T\}$, the alternating minimization (AM) framework \cite{opt} is appropriate to be exploited for the algorithm design.

\subsection{Optimization via Alternating Minimization (AM)}
\label{sec:formulation_2}
Following the principle of AM, the decoupled variables $\mathbf{w}$ and $\{\phi_t\}^T_{t=1}$ are updated alternately. More specifically, given the initialization $\mathbf{w}^0$ and $\{\phi^0_t\}^T_{t=1}$, the alternating update rule consists of the following two steps.

\noindent{\it 1) \underline{Optimizing $\{\phi_t\}^T_{t=1}$:}} In the iteration $\tau+1$, variable $\mathbf{w}$ is fixed to its latest value $\mathbf{w}^{\tau}$, and thus problem \eqref{EQ:14} reduces to \begin{align}
    \max_{{\{\phi_t\}}^{T}_{t=1}} \sum_{t=1}^{T} \sqrt{\eta(t)}  \Re\{\mathbf{v}_{m_0(t)}^{H} \mathbf{Q}(t)^H {\mathbf{w}^{\tau}}\exp(j\phi_t)\}. \label{opt_phi}
\end{align}
Appendix \ref{proof:equivalence} shows that the optimal $\phi^{\tau+1}_t$ should satisfy 
\begin{align}
      \Re\{\mathbf{v}_{m_0(t)}^{H} \mathbf{Q}(t)^H {\mathbf{w}^{\tau}}\exp(j\phi^{\tau+1}_t)\} = |\mathbf{v}_{m_0(t)}^{H} \mathbf{Q}(t)^H {\mathbf{w}^{\tau}}|,
\end{align}
which implies that 
\begin{align}
\exp(j\phi^{\tau+1}_t) =  \frac{ \overline{(\mathbf{v}_{m_0(t)}^{H} \mathbf{Q}(t)^H \mathbf{w}^{\tau})}}{|\mathbf{v}_{m_0(t)}^{H} \mathbf{Q}(t)^H \mathbf{w}^{\tau}|}. \label{exp_express}
\end{align} 
In particular, $\phi^{\tau+1}_t$ can be specified as 
\begin{align}
    \phi^{\tau+1}_t = \arctan{\frac{b_t^\tau}{a_t^\tau}},\label{15}
\end{align}
where 
\begin{align}
    a^\tau_t = \Re\left\{\frac{{\mathbf{v}_{m_0(t)}^{H} \mathbf{Q}(t)^H \mathbf{w}^{\tau}}}{|\mathbf{v}_{m_0(t)}^{H} \mathbf{Q}(t)^H \mathbf{w}^{\tau}|}\right\};\\
    b^\tau_t = - \Im\left\{\frac{{\mathbf{v}_{m_0(t)}^{H} \mathbf{Q}(t)^H \mathbf{w}^{\tau}}}{|\mathbf{v}_{m_0(t)}^{H} \mathbf{Q}(t)^H \mathbf{w}^{\tau}|}\right\}.
\end{align}

 \noindent{\it 2) \underline{Optimizing $\mathbf{w}$:}} Similarly, after fixing $\{\phi^{\tau+1}_t\}^{T}_{t=1}$, problem \eqref{EQ:14} can be reorganized as 
   \begin{align}
     \min_{\mathbf{w}} {\mathbf{w}}^H \mathbf{A} \mathbf{w} - 2\Re\{{(\mathbf{b}^{\tau+1})}^H \mathbf{w}\},\label{opt_w}
   \end{align}
   where 
   \begin{align}
     \mathbf{A} &\triangleq \lambda \mathbf{I}  + \sum_{t=1}^{T} \mathbf{Q}(t)\mathbf{v}_{m_0(t)}  \mathbf{v}_{m_0(t)}^H \mathbf{Q}(t)^H,\label{I_A} \\ 
     \mathbf{b}^{\tau+1} &\triangleq \sum_{t=1}^{T} \sqrt{\eta(t)}(\overline{\exp(j\phi^{\tau+1}_t)})\mathbf{Q}(t)\mathbf{v}_{m_0(t)} .\label{I_b}
   \end{align}
By taking the derivative of the objective function in problem \eqref{opt_w}, it can be shown the closed-form solution of problem \eqref{opt_w} is \cite{opt}:
\begin{align}
    \mathbf{w} = \mathbf{A}^{-1} \mathbf b^{\tau+1}, \label{19}
\end{align}
which will be assigned to $\mathbf{w} ^{\tau+1}$.

Based on the alternating update rule consisting of \eqref{15} and \eqref{19}, the AM-based beamforming vector estimation algorithm is summarized as \textbf{Algorithm \ref{alg:1}}.

\begin{algorithm}[!tbp]
\caption{\bf\hspace{-0.1cm}: AM-based Beamforming Vector Estimation}

\noindent {\bf Input}: weighting matrix $\{\mathbf {Q}(t)\}^T_{t=1}$, regularization parameter $\lambda>0$.

$~~$\noindent {\bf Initialization}: ${\mathbf w}^{0}$, $\{\phi^0_t\}^T_{t=1}$.

$~~$For the iteration $\tau+1 (\tau \geq 0)$:

$~~~~~~$\noindent Update $ \{\phi^{\tau+1}_t\}^{T}_{t=1}$ via equation \eqref{15};

$~~~~~~$\noindent Update $ {\mathbf w}^{\tau+1}$  via equation \eqref{19};

$~~$\noindent {\bf{Until Convergence}}.

\noindent {\bf Output}: beamforming vector $\mathbf{w}_T \triangleq \frac{\mathbf{w}^{\tau+1}}{\|\mathbf{w}^{\tau+1}\|}$. 
\label{alg:1}
\end{algorithm}

\noindent\emph{Remark 1 (Convergence Property)}: 
In particular, since the proposed \textbf{Algorithm \ref{alg:1}} exploits the AM framework, its convergence is guaranteed by the convergence theory of AM framework \cite[page 324]{opt}. In brief, the convergence theory states that for a particular problem with several decoupling variables, if objective functions of its subproblems are monotonically decreasing (or increasing) and have a corresponding unique minimum (or maximum), the AM based algorithm can converge to the stationary point. It is easy to check that problems \eqref{opt_phi} and \eqref{opt_w} satisfy the above conditions, and thus \textbf{Algorithm 1} can converge according to the convergence theory of AM framework.

\subsection{Learning Regularization Parameter $\lambda$ via Empirical Bayes}
\label{sec:bayes_regularization}
 In problem \eqref{12}, the regularization parameter $\lambda$ balances the shrinkage effect on $\mathbf w$ and the CQI data fit. Hence, the choice of regularization parameter $\lambda$ would affect the performance of beamforming vector estimation in \textbf{Algorithm 1}. As $\lambda$ is tied to the unknown $\hat{\sigma}_1$, manual parameter tuning becomes both time-consuming and prone to inconsistencies in acquiring the parameter value. Instead, our proposal involves learning $\lambda$ from the received feedback data using the empirical Bayes technique. This technique leverages the relationship between the optimization problem \eqref{EQ:14} and a specific Bayesian model (see \textbf{Bayesian Model}). Such a relation is revealed by the following proposition.

\begin{proposition}
     Given $\{\phi_t = \hat{\phi}_t\}_{t=1}^T$,  optimization problem \eqref{EQ:14} is equivalent to the maximum a posterior estimation (MAP) problem for \textbf{Bayesian Model} when $\lambda = \frac{\alpha}{\beta}$, where $\alpha$ and $\beta$ are the precision parameters of the likelihood \eqref{likelihood} and prior \eqref{prior}, respectively. 
    \proof see Appendix \ref{proof:proposition2}.
    \label{pro:2}
\end{proposition}

\setlength{\textfloatsep}{10pt}
\begin{algorithm}[!t]
\caption*{\bf Bayesian Model \hspace{-0.1cm}: Equivalent Bayesian Linear Regression Model}
    \vspace{0.2cm}
    \textbf{Likelihood:}
   \begin{align}
    p(\mathbf y|\mathbf{w} ; \beta) \propto \exp\{-\frac{\beta}{2} \|\mathbf{y}-\mathbf{\Phi}\mathbf{w} \|^2\}, \label{likelihood}
  \end{align}
  where 
    \begin{align}
      \mathbf{\Phi} &=\left[\begin{array}{c} \exp(j\hat{\phi}_1)\mathbf{v}_{m_0(1)}^{H} \mathbf{Q}(1)^H \\ \exp(j\hat{\phi}_2) \mathbf{v}_{m_0(2)}^{H} \mathbf{Q}(2)^H \\ \vdots \\ \exp(j\hat{\phi}_T) \mathbf{v}_{m_0(T)}^{H} \mathbf{Q}(T)^H\end{array}\right],\label{likelihood_PHI}\\
      \mathbf{y} &= [\sqrt{\eta(1)},\sqrt{\eta(2)},\cdots,\sqrt{\eta(T)}]^T.
    \end{align} 

\vspace{0.1cm}
\textbf{Prior Distribution:}
\vspace{-0.1cm}
\begin{align}
    \mathbf{w}  &\sim  \mathcal{N}\left(\mathbf{0}, \alpha^{-1} \mathbf{I}_{N_A}\right). \label{prior}
\end{align}
\end{algorithm}

Notice that the choice of $\{\exp(j \hat{\phi}_t)\}_{t=1}^T$ directly impacts the values of $\alpha$, $\beta$, and consequently, the performance of $\lambda$. The optimal choice for $\{\exp(j \hat{\phi}_t)\}_{t=1}^T$ is to set them as the ground-truth values, i.e., $\exp(j \hat{\phi}_t) = \frac{ \overline{(\mathbf{v}_{m_0(t)}^{H} \mathbf{Q}(t)^H \mathbf{w}^\star)}}{|\mathbf{v}_{m_0(t)}^{H} \mathbf{Q}(t)^H \mathbf{w}^|}$. However, such a choice is unachievable due to the unknown $\mathbf{w}^\star$. Therefore, an alternative scheme is to approximate $\mathbf{w}^\star$ with its historical estimation, e.g., $\mathbf{w}_{T-1}$.

Furthermore, \textbf{Proposition \ref{pro:2}} indicates the optimal $\lambda$ can be estimated as $\frac{\alpha}{\beta}$ when the optimal value of $\{ \alpha, \beta \}$ in \textbf{Bayesian Model} has been estimated via an empirical Bayes approach \cite[page 596-604]{ml}. Based on the marginal distribution:
\begin{align}
    & p(\mathbf{y} | \alpha, \beta) = \int p(\mathbf{y}|\mathbf{w}, \beta) p(\mathbf{w}| \alpha) \mathrm{d} \mathbf{ w} \nonumber\\
    &= \left(\frac{\alpha}{2\pi}\right)^{\frac{N_A}{2}}\left(\frac{\beta}{2\pi}\right)^{\frac{T}{2}}\int \exp\{-V(\mathbf{ w})\}\mathrm{d} \mathbf{ w},\label{25}
\end{align}
where 
\begin{align}
    V\left(\mathbf{w}\right) &= \frac{\beta}{2}\|\mathbf{y}-\mathbf{\Phi} \mathbf{ w}\|^{2}+\frac{\alpha}{2} \mathbf{w}^{H} \mathbf{w},\label{26}
\end{align}
the optimal $\{\alpha,\beta\}$ estimate can be acquired by solving the following evidence maximization problem \cite{ml}:
\begin{align}
  &~~~\max_{\alpha,\beta} ~~\ln p(\mathbf{y}|\alpha, \beta)\nonumber\\
  &=\max_{\alpha,\beta}~~\frac{N_A}{2} \ln \alpha+\frac{T}{2} \ln \beta-E\left(\mathbf{m} \right) -\frac{1}{2} \ln |\mathbf{B}|-\frac{T}{2} \ln (2 \pi), \label{27}
\end{align}
where 
\begin{align}
  \mathbf{B}&=\alpha \mathbf{I}+\beta \mathbf{\Phi}^{H} \mathbf{\Phi}, \label{28}\\
  \mathbf{m}&=\beta \mathbf{B}^{-1} \mathbf{\Phi}^{H} \mathbf{y},  \label{29} \\
  E\left(\mathbf{m} \right) &= \frac{\beta}{2}\|\mathbf{y}-\mathbf{\Phi} \mathbf{m}\|^{2}+\frac{\alpha}{2} \mathbf{m}^{H} \mathbf{m}.\label{30}
\end{align}

In particular, \cite[page 167]{bishop} has shown that the fixed-point method\cite{fix} method can solve problem \eqref{27} well, which is summarized as \textbf{Algorithm \ref{alg:2}}. Hence, the optimal regularization parameter $\lambda$ can be estimated from \textbf{Algorithm 
\ref{alg:2}}.
\begin{algorithm}[!tbp]
\label{alg:2}
\caption{\bf\hspace{-0.1cm}: Empirical Bayes based Regularization Parameter Learning}

\noindent {\bf Input}: Matrix $\mathbf{\Phi}$ and vector $\mathbf{y}$.

$~~$\noindent {\bf Initialization}: $\alpha^{0}, \beta^{0}$.

$~~$For the iteration $k ~(k \geq 1)$:

$~~~~~~$\noindent Compute the eigenvalues $\mu_1,\cdots,\mu_{N_A}$ of $\beta^{k-1} \boldsymbol{\Phi}^{H} \boldsymbol{\Phi}$;
\vspace{0.15cm}

$~~~~~~$\noindent Compute $ \gamma $ via equation $\gamma=\sum_{i = 1}^{N_A} \frac{\mu_{i}}{\alpha^{k-1}+\mu_{i}}$;
\vspace{0.15cm}

$~~~~~~$\noindent Update $ \alpha^{k} $ via equation $\alpha^k=\frac{\gamma}{\mathbf{m}^{H} \mathbf{m}}$;
\vspace{0.15cm}

$~~~~~~~~$\noindent Update $ \beta^{k}$ via equation $\frac{1}{\beta^k}=\frac{1}{T-\gamma} \|\mathbf{y}-\mathbf{\Phi} \mathbf{m}\|^{2}$;

$~~$\noindent \textbf{Until Convergence}.

\noindent {\bf Output}: regularization parameter $\lambda=\frac{\alpha^{k}}{\beta^{k}}$. 
\label{alg:2}
\end{algorithm}
  
\subsection{The Extension to Multi-Stream Beamforming Scenario}
\label{sec:multi-stream}

Section \ref{sec:formulation} outlines a method designed for acquiring a beamforming vector, tailored for enhancing the directional transmission of a single data stream. However, current wireless communication systems widely adopt concurrent directional transmission of multiple data streams to bolster transmission efficiency. In the multi-stream scenario, multiple data streams are required to be assigned to distinct beamforming vectors, while ensuring that these vectors are orthogonal to prevent interference among the data streams. Consequently, in the multi-stream context, the task shifts from estimating a single beamforming vector to the simultaneous estimation of several orthogonal beamforming vectors. Such a task becomes even more intricate due to the interrelatedness among these vectors. To tackle this challenge, we propose a sequential estimation approach that leverages previously obtained estimates to construct orthogonal constraints for subsequent estimations.

Based on the multi-stream relationship presented in Section \ref{sec:multi-stream link} and single-stream problem formulation in Section \ref{sec:formulation_1}, the multi-stream beamforming estimation problem is formulated as follows. 
\begin{align}
       &\min_{\{\mathbf{w}_k\}_{k=1}^{N_U}} \sum_{k=1}^{N_U}\sum_{t=1}^{T} \left(|\mathbf{v}_{k,m_0(t)}^{H} \mathbf{Q}(t)^H\mathbf{w}_k| - \sqrt{ \eta_{k}(t)} \right)^2+\lambda_{k}||\mathbf w_k||^2 ~~\nonumber \\ 
    & ~~~~~~\mathrm{s.t.} ~~~~~\mathbf w^H_k   \mathbf{w}_{k^\prime}  = \mathbf{0}, ~~~~~\forall 1\leq k\neq k^\prime \leq N_U,\label{subprob_multi_stream_0}
\end{align}
where $\mathbf{w}_k$ corresponds to $k$-th beamforming vector. Here, the orthogonal constraints are introduced since beamforming vectors $\{\mathbf{w}^\star_k\}_{k=1}^{N_U}$ are orthogonal. However, these orthogonal constraints couple variables, posing a challenge in solving problem \eqref{subprob_multi_stream_0}. To tackle this challenge, we propose to decompose problem \eqref{subprob_multi_stream_0} as $N_U$ subproblems and solve them sequentially. To be specific, 
the $k$-th subproblem is given by
\begin{align}
    \mathrm{P}_k: ~~ &\min_{\mathbf w_k} \sum_{t=1}^{T} \left(|\mathbf{v}_{k,m_0(t)}^{H} \mathbf{Q}(t)^H\mathbf{w}_k| - \sqrt{ \eta_{k}(t)} \right)^2+\lambda_{k}||\mathbf w_k||^2 ~~\nonumber \\ 
    & ~~~~~~\mathrm{s.t.} ~~\mathbf w^H_k \mathbf{\hat W}_{k-1} = \mathbf{0}, \label{subprob_multi_stream_1}
\end{align}
where $\mathbf{\hat W}_{k-1} \triangleq [\hat{\mathbf{w}}_1,\hat{\mathbf{w}}_2,\cdots,\hat{\mathbf{w}}_{k-1}]$ and $\mathbf{\hat{w}}_i$ represents the solution of $i$-th subproblem $\mathrm{P}_i$.  
Notice that subproblem $\mathrm{P}_1$ has no orthogonal constraints.  By sequentially solving these subproblems in ascending order of their indices, the complicated synchronous estimation of multiple beamforming vectors is simplified into several asynchronous beamforming vector estimations. Furthermore, problem \eqref{subprob_multi_stream_1} can be reorganized as an unconstrained problem by using variable substitution $\mathbf w_k = \mathbf{B}_k \mathbf u_k$ as follows:
\begin{align}
  \mathrm{P}_k: ~~&\min_{\mathbf u_k} \sum_{t=1}^{T} \left(|\mathbf{v}_{k,m_0(t)}^{H} \mathbf{Q}(t)^H\mathbf{B}_k \mathbf u_k| \!-\! \sqrt{ \eta_{k}(t)} \right)^2\nonumber\\
  &~~~~~~~~~~~~~~~~~~~~~~~~~~~~~~~~~~~~~~~~~+\lambda_{k}||\mathbf u_k||^2,
   \label{subprob_multi_stream_3}
\end{align}
where $\mathbf u_k \in \mathbb{C}^{N_A-k+1}$ and $\mathbf{B}_k \in \mathbb{C}^{N_A \times (N_A-k+1)}$ is the basis matrix\footnote{For a given space, its basis matrix is composed of all the independent unit-norm basis of this space.} of $\mathrm{null}(\mathbf{\hat W}^H_{k-1})$ satisfying
\begin{align}
    &\mathbf{B}^H_k\mathbf{\hat W}_{k-1}= \mathbf{0}; \label{B_1}\\
    &\mathbf{B}^H_k\mathbf{B}_k = \mathbf{I}. \label{B_2}
\end{align}
Particularly, problem \eqref{subprob_multi_stream_3} can be solved via {\bf Algorithm \ref{alg:1}} by substituting the input parameter $\mathbf{Q}(t)$ with $\mathbf{B}_k^H\mathbf{Q}(t)$.  Meanwhile, based on the same substitution, the regularization parameter $\lambda_k$ can be auto-tuned via {\bf Algorithm \ref{alg:2}}. The multi-stream beamforming algorithm sequentially solves problems $\{\mathrm{P}_k\}_{k=1}^{N_U}$, summarized as \textbf{Algorithm \ref{alg:3}}.

\begin{algorithm}[H]
    \caption{\bf\hspace{-0.1cm}: Multi-Stream Beamforming Vectors Estimation Algorithm with CQIs}
  
  \noindent {\bf Input}: Weighting matrix $\{\mathbf {Q}(t)\}_{t=1}^T$, CQI $\{\eta (t)\}_{t=1}^T$.
  
  $~~$\noindent {\bf Initialization}: $\{\mathbf w^{0}_k\}^{N_U}_{k=1}$, $\{\mathbf \eta_{k} (t)\}^{N_U,T}_{k=1,t=1}$;
  
  $~~$For the index $k = 1,\cdots, N_U$:
  
  $~~~~~~$If $k = 1$: 

  $~~~~~~~~~$ $\mathbf{B}_k$ is set as $\mathbf{I}$;

  $~~~~~~$If $k\geq 2$: 

  $~~~~~~~~~$Use $\{\mathbf{w}_i\}^{k-1}_{i=1}$ to compute $\mathbf{B}_k$ via \eqref{B_1} and \eqref{B_2};

  $~~~~~~$ Calculate  $\mathbf u_k$ by solving problem \eqref{subprob_multi_stream_3} with \textbf{Algorithm \ref{alg:1}} 
  
  $~~~~~~$and \textbf{Algorithm \ref{alg:2}};
  
  $~~~~~$ Compute $\mathbf{w}_k$ via $\mathbf{w}_k = \mathbf{B}_k\mathbf u_k$.

  \noindent {\bf Output}: unit-norm vectors $\{\mathbf{w}_k\}^{N_U}_{k=1}$.
  \label{alg:3}
  \end{algorithm}

\section{Numerical Results}
\label{sec:numerical}

 In this section, numerical results are presented to showcase the performance of the proposed algorithms. In particular, the real-life system independently configures beamforming vectors for each UE. Therefore, to emulate practical configurations, we conduct numerical experiments based on the FDD communication system consisting of a single UE with $N_U = 2$ antennas and a BS with $N_A = 32$ antennas and $N_P = 8$ ports. The QuaDriGa platform is utilized to simulate the communication system with the center frequency of the downlink channel set to $1.275$ GHZ. Based on the simulated communication system, the channel covariance matrices $\mathbf{C}$ are computed via the generated channel samples. Type I codebook, as specified in 5G NR standard  \cite{code1,22}, is used in the limited feedback scheme. To ensure a fair comparison, the following algorithms have the same feedback bit quantification. Specifically, at each communication round, 36 feedback bits are exploited, including 4 bits for PMI and 32 bits for CQI. Additionally, the pilot weighting matrices $\{\mathbf{Q}(t)\}_{t=1}^{T}$ are semi-unitary and randomly sampled from a Gaussian distribution except for $\mathbf{Q}(1)$, which is selected from a pre-defined beam pattern weighting set\footnote{In beam management, the BS initiates an initial broad beam sweep by employing pre-defined beam patterns, resulting in the generation of $\mathbf{Q}(1)$. Following this, the BS conducts a more precise beam sweep using CSI-RS, leading to the generation of subsequent $\mathbf{Q}$ matrices\cite[6.1.6.1]{code1}.}. 
 
The initialization for each algorithm is respectively shown below. For \textbf{Algorithm \ref{alg:1}}, we set $\phi_t^0 = 0, t=1,2,\cdots,T$, and ${\mathbf w}^0 \sim \mathcal{CN}(\mathbf 0, \mathbf I_{N_A})$. For \textbf{Algorithm \ref{alg:2}}, the initial parameters $\alpha^0$ and $\beta^0$ are randomly generated from $[0, 1]$. In \textbf{Algorithm \ref{alg:3}}, there are ${\mathbf w}^0_k \sim \mathcal{CN}(\mathbf 0, \mathbf I_{N_A}), k=1,\cdots,N_U$, and $\{\mathbf \eta_{k} (t)\}^{N_U,T}_{k=1,t=1}$ generated from the average partition scheme (see Section \ref{sec:multi-stream link}). In addition, all the numerical results are obtained by averaging over $100$ simulation trials.
 
The beam estimation performance is measured by the beam precision, which is derived from the SNR formulation (see \eqref{snr}) and defined as 
\begin{align}
    \frac{\mathrm{Tr}(\mathbf{W}^H{\mathbf C}\mathbf{W})}{\mathrm{Tr}(\mathbf{C})},
\end{align}
where $\mathbf{W}$ represents the estimated beamforming vectors and $\mathbf{C}$ is the ground-truth DCCM. In particular, the beamforming matrix will be reduced to a beamforming vector in the single-stream scenario. Correspondingly, the beam precision is reduced to
\begin{align}
    \frac{\mathbf{w}^H{\mathbf C}\mathbf{w}}{d},
\end{align}
where $d$ is the largest eigenvalue of $\mathbf C$. 

\begin{figure}[!t]
    \centering                                                                                                          
    \includegraphics[width= 3.5in]{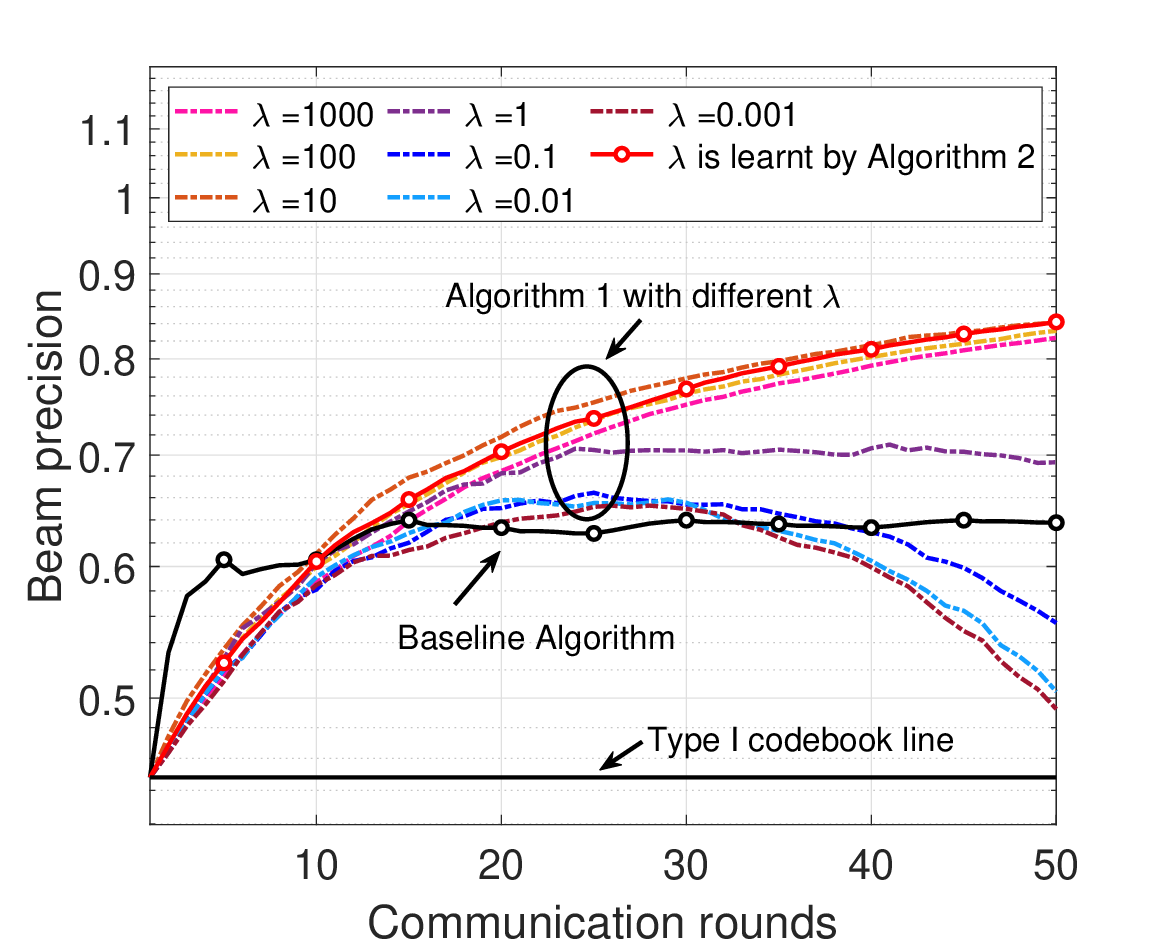}
    \caption{The beam precisions versus communication rounds (single-stream).}
    \label{p1}
\end{figure}

The superiority of the Bayesian-based auto-tuning mechanism (i.e., {\bf Algorithm \ref{alg:2}}) is demonstrated in Figure \ref{p1}. Based on the proposed single-stream beamforming estimation algorithm (i.e., {\bf Algorithm \ref{alg:1}}), the Bayesian-based auto-tuning mechanism gives the best performance among all the constant regularization parameters schemes except for $\lambda = 10$. In particular, the ill-suited regularization parameters will degrade the performance, such as $\lambda = 0.1,0.01,0.001$. This comparison highlights the superiority of the proposed Bayesian-based auto-tuning mechanism in estimating the optimal beamforming vector. It is worth mentioning that the performance of our proposed Bayesian-based auto-tuning mechanism depends on the configuration of $\{\hat{\phi}_t\}_{t=1}^T$, as discussed in Section \ref{sec:bayes_regularization}. Since we use an estimation of $\mathbf{w}^\star$, rather than $\mathbf{w}^\star$ itself, to determine $\{\hat{\phi}_t\}_{t=1}^T$, the proposed mechanism may not always yield the optimal penalty parameter. This explains why the setting $\lambda = 10$ outperforms our proposed scheme.

Meanwhile, Figure \ref{p1} shows the comparison between the proposed auto-tuning mechanism based beamforming estimation algorithm and the baseline algorithm in \cite{patent} (labeled as baseline algorithm). The baseline algorithm is widely used by a leading global provider of information and communications technology infrastructure in typical massive MIMO 5G systems, and thus being provided as the benchmark in this paper. In particular, the baseline algorithm estimates the beamforming vectors at communication round $t$ by solving
\begin{align}
    \max_{\mathbf{W}^H\mathbf{W} = \mathbf{I}}\mathrm{Tr}\left(\mathbf{W}^H\left(\frac{1}{t}\sum_{i=1}^t \eta(i) \mathbf{Q}(i)\mathbf{v}_{m_0(i)} \mathbf{v}^H_{m_0(i)}\mathbf{Q}(i)^H\right)\mathbf{W}\right),
\end{align}
where $\mathbf{Q}(i) = \mathbf{Q}(1)\mathbf{O}_i$ and $\mathbf{O}_i$ is a unitary matrix\footnote{ $\mathbf{O}_i$ is chosen from a mutually unbiased bases (MUB) codebook, specified in the patent \cite{patent}( https://patents.google.com/patent/WO2017206527A1/en).}. In the single-stream scenario, matrix $\mathbf{W}$ is reduced to a vector $\mathbf{w}$. In addition, the beam precision obtained by using only the Type I codebook directly is also taken as another benchmark to evaluate the performance gains from the proposed algorithm. It can be observed that the proposed algorithm outperforms the benchmarking algorithm \cite{patent} after $10$ communication rounds. Meanwhile, as the number of communication rounds increases, the beam precision of the proposed algorithm continues to improve. These results show the effectiveness of the proposed algorithm in enhancing beamforming performance.

\begin{figure}[!t]
    \centering                                                                                         
    \includegraphics[width= 3.5in]{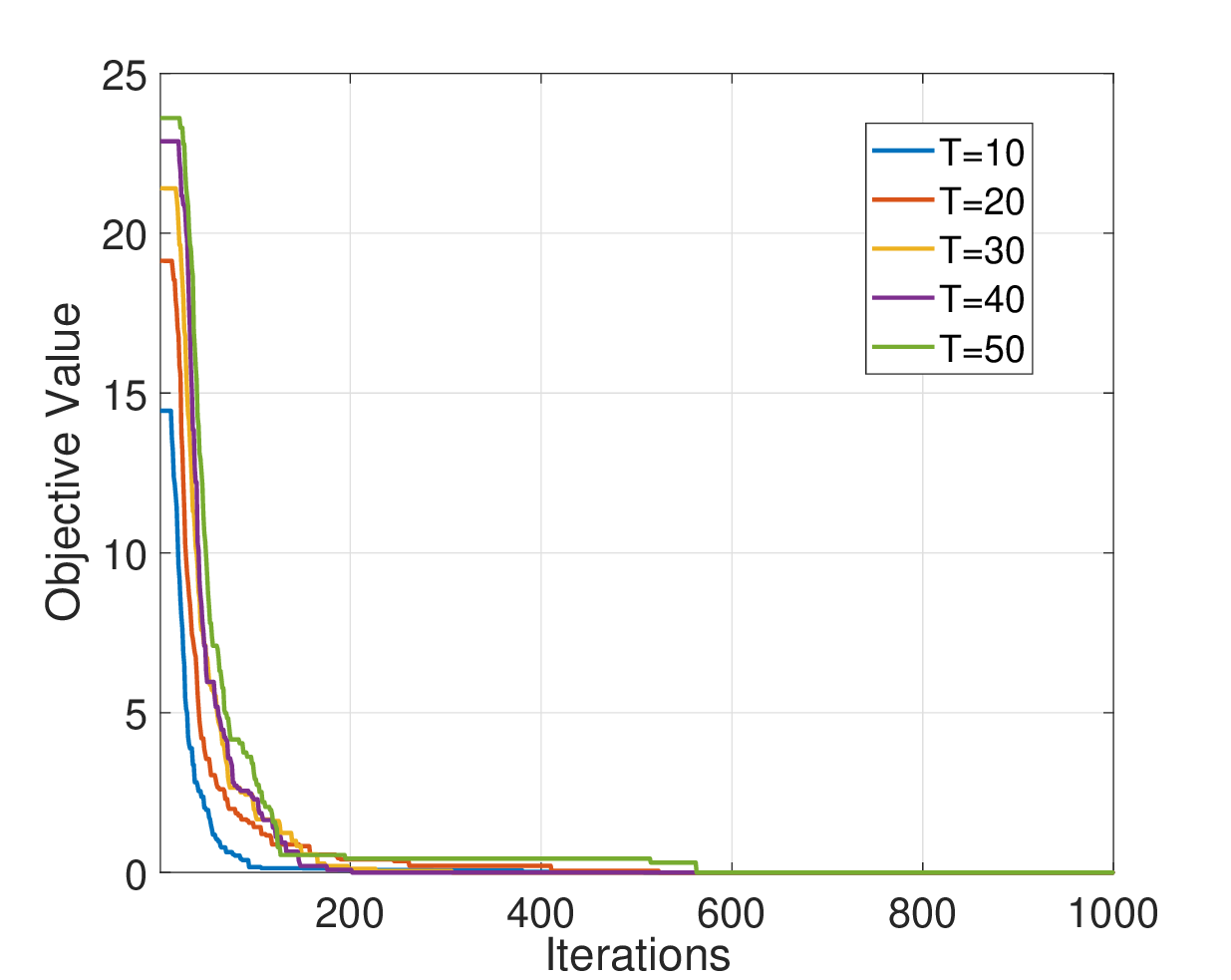}
    \caption{The objective values associated with {\bf Algorithm \ref{alg:1}} at different values of $T$ versus iterations.}
    \label{conv_p1}
\end{figure}

Figure \ref{conv_p1} illustrates the convergence of {\bf Algorithm \ref{alg:1}}. The algorithm is employed to address problem \eqref{12} at different given values of $T$. Regardless of the value of $T$, the objective value associated with the algorithm decreases with each iteration, and remains almost constant after approximately $200$ iterations. This phenomenon indicates that the alternating update of variables $\mathbf{w}$ and $\{\phi_t\}_{t=1}^T$ could stop after $200$ iterations, i.e., the algorithm converges to a stationary point.

\begin{figure}[!t]
    \centering
    \includegraphics[width= 3.6in]{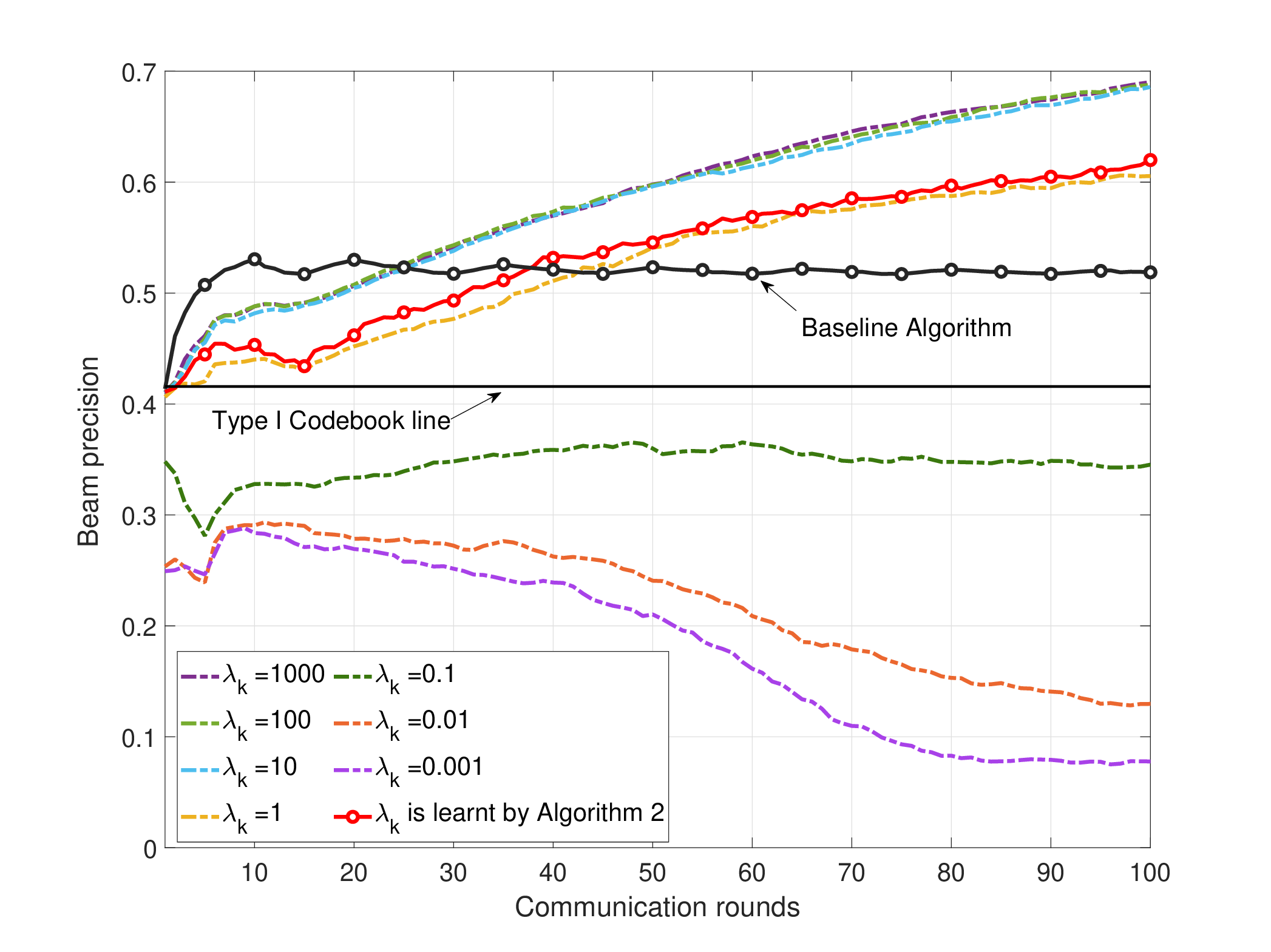}
    \caption{The beam precision versus communication rounds when the accurate partition of CQI is unknown (multi-stream).}
    \label{p4}
\end{figure}

Similarly, Figure \ref{p4} illustrates the performance of {\bf Algorithm \ref{alg:3}} with different parameter setting schemes. In the multi-stream scenario, unlike the single-stream case, the auto-tuning mechanism does not outperform constant parameter schemes. In particular, the constant parameter scheme with a fixed regularization parameter of $10,100,1000$ brings more performance gain for the proposed algorithm than the auto-tuning scheme. However, the proposed auto-tuning mechanism can make the proposed beamforming estimation algorithm avoid poor performance caused by inappropriate regularization parameters (e.g., $0.1,0.01,0.001$), which do not even surpass the benchmark line (i.e., Type I Codebook line). Furthermore, the proposed algorithm still effectively utilizes the information of CQIs for performance enhancement in the multi-stream case. However, the proposed auto-tuning mechanism based algorithm requires more communication rounds to surpass the baseline algorithm compared to the single-stream case.

\begin{figure}[!t]
 
    \centering
    \includegraphics[width= 3.5in]{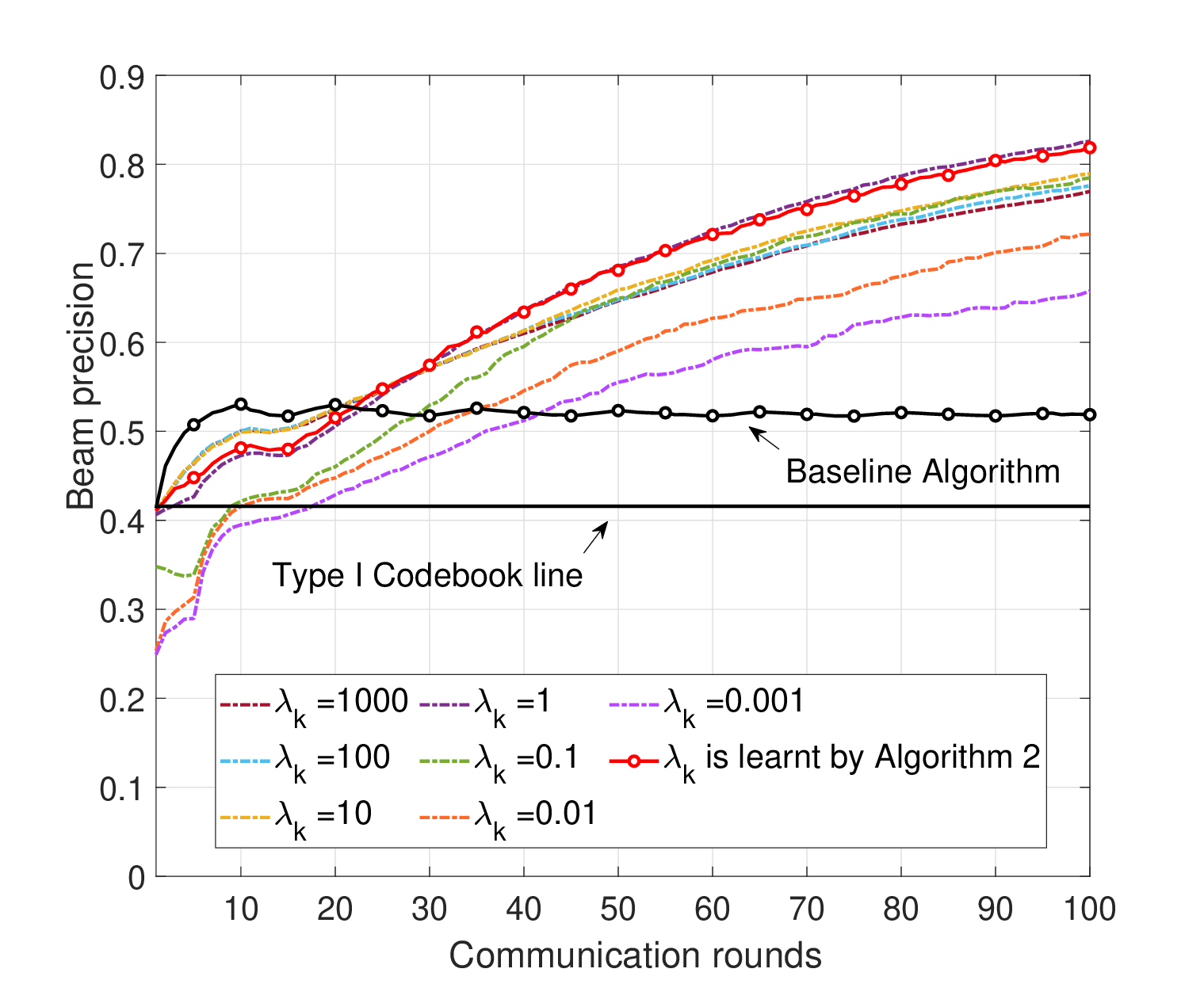}
    \caption{The beam precision versus communication rounds when the accurate partition of CQI is given (multi-stream).}
    \label{p5}
\end{figure}

Notice that the inaccurate partition of multi-stream CQI results in poor performance of the proposed algorithm shown in Figure \ref{p4}. Each component of the CQI partition serves as a reference for estimating one eigenvector of DCCM $\mathbf{C}$. Therefore, an imprecise CQI partition directly results in poor performance, especially when the regularization effect is weak (i.e., when the regularization parameter is small such as 0.1, 0.01, and 0.001 in Figure  \ref{p4}). Assuming the accurate separation of multi-stream CQI is known, the performance of the proposed multi-stream beamforming estimation algorithm with different regularization parameters, including the proposed auto-tuning mechanism, is shown in Figure \ref{p5}. Under this setting, the auto-tuning mechanism can provide the more appropriate regularization parameter, which leads to better beam precision than the constant parameter schemes. Specifically, the auto-tuning mechanism surpasses other regularization parameter settings after about $22$ communication rounds. Meanwhile, the auto-tuning based algorithm with the accurate separation of multi-stream CQI assumption surpasses the benchmark algorithm at about the $22$-th communication round. In contrast, the algorithm without accurate separation of multi-stream CQI requires more communication rounds to surpass the benchmark, as shown in Figure \ref{p4}.

 Furthermore, although the proposed algorithms demonstrate performance enhancements as illustrated in Figure \ref{p1}, \ref{p4} ,\ref{p5}, they cannot achieve unlimited gains in beam precision, since beam precision is bounded by a maximum value of 1. Meanwhile. the proposed algorithms can not reach the upper bound due to estimation errors, which stem from quantization errors in the codebook, approximation errors between CQI and optimal beamforming vectors, and CQI partition errors. The MIMO dimension plays a pivotal role in determining algorithm performance as well. A larger MIMO dimension results in a higher-dimensional variable, necessitating the algorithm to estimate more parameters. Specifically, as depicted in Figure \ref{p1}, the performance of our proposed algorithm exhibits notable variations across different antenna configurations. With an increase in the number of antennas, a decline in performance at each communication round becomes evident. To mitigate this decline, additional feedback bits or communication rounds may be necessary.
\begin{figure}[htbp]
    \centering                                                                                         
    \includegraphics[width= 3.4in]{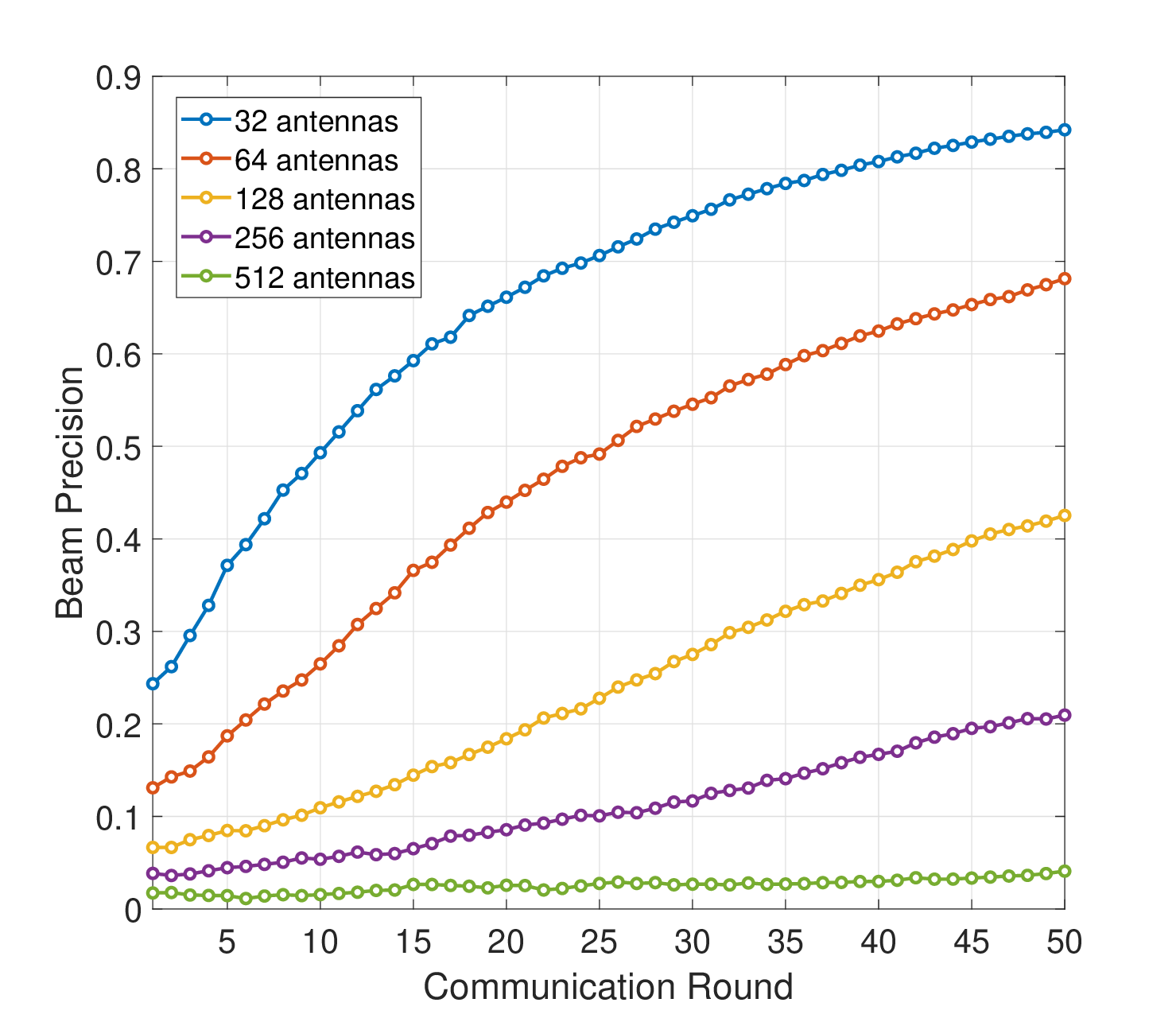}
    \caption{The beam precisions of {\bf Algorithm \ref{alg:1}} with different antenna configurations.}
    \label{p_dim}
\end{figure}

\section{Conclusion}
\label{sec:conclusion}

In this paper, we focus on the beamforming problem in  5G NR FDD massive MIMO systems. In particular, we present novel beamforming algorithms for both single-stream and multi-stream scenarios.  Specifically, to enhance the performance of beamforming vectors, we have leveraged the information contained in CQIs to propose a principled beamforming estimation problem (see Section \ref{sec:formulation_1} and Section \ref{sec:multi-stream}). By bridging the proposed problem formulation and the Bayesian model, we propose a Bayesian learning-based regularization scheme. By exploiting such a scheme, our proposed algorithm can outperform the widely used approach and enable better beamforming estimates as communication rounds increase. Notably, this improved performance can be observed in both single-stream and multi-stream scenarios, showcasing the potential of leveraging CQIs to enhance the performance of massive MIMO systems.  

However, in the multi-stream scenario, the accuracy of the CQI partition significantly affects the performance of our proposed algorithm. Unfortunately, the limited feedback scheme does not provide any multi-stream CQI partition information, resulting in no guarantee of the accuracy of the CQI partition. Therefore, accurate CQI partition acquisition is a promising future research direction. 

\section*{Acknowledgments}
We would like to express our sincere gratitude to Congliang Chen for his valuable suggestions, insightful discussions, and meticulous proofreading. 
\appendix

\subsection{The Proof of {\bf Proposition \ref{prop:1}}}
\label{proof:proposition1}
Given the assumption $\mathbf{w}^\star\in\mathrm{Col}(\mathbf{Q}(t))$, there are the following lemmas.
\begin{lemma}

Given a semi-unitary matrix $\mathbf{Q}(t)$, $\mathbf{w}^\star\in\mathrm{Col}(\mathbf{Q}(t))$, and a unit-norm vector $\mathbf{u}$ satisfying $\mathbf{u}^H\mathbf{w}^\star = 0$, there are
\begin{align}
    \mathbf{{w}^\star}^H\mathbf{Q}(t)\mathbf{Q}(t)^H\mathbf{w}^\star &= 1;\label{prop1_lemma_1_2} \\
    \mathbf{u}^H\mathbf{Q}(t)\mathbf{Q}(t)^H\mathbf{w}^\star &= 0; \label{prop1_lemma_1_1} \\
     \mathbf{u}^H\mathbf{Q}(t)\mathbf{Q}(t)^H\mathbf{u} & \leq 1.\label{prop1_lemma_1_3}
\end{align}
\begin{proof}

Due to $\mathbf{w}^\star\in\mathrm{Col}(\mathbf{Q}(t))$, the projection of $\mathbf{w}^\star$ on space $\mathrm{Col}(\mathbf{Q}(t))$ is still itself, i.e.,  
\begin{align}
    \mathbf{w}^\star= \mathbf{Q}(t)(\mathbf{Q}(t)^H\mathbf{Q}(t))^{-1}\mathbf{Q}(t)^H\mathbf{w}^\star = \mathbf{Q}(t)\mathbf{Q}(t)^H\mathbf{w}^\star. \label{prop1_1}
\end{align}
Hence, from \eqref{prop1_1}, we can conclude that 
\begin{align}
    &{\mathbf{w}^\star}^H\mathbf{Q}(t)\mathbf{Q}(t)^H\mathbf{w}^\star =  {\mathbf{w}^\star}^H\mathbf{w}^\star = 1;\\
    &\mathbf{u}^H \mathbf{Q}(t)\mathbf{Q}(t)^H\mathbf{w}^\star = \mathbf{u}^H  \mathbf{w}^\star =0.
\end{align}
Moreover, the inequality \eqref{prop1_lemma_1_3} is trivial. 
\end{proof}
\label{lemma:1}
\end{lemma}

From the definition of PMI in \eqref{PMI} and the eigenvalue decomposition of DCCM $\mathbf{C}$, it can be seen that 
\begin{align}
      m_0(t) &= \arg\max_{m=1,\cdots,M} \mathbf v^H_m  \mathbf Q(t)^H \mathbf{C} \mathbf Q(t)  \mathbf v_m \nonumber\\
      &= \arg\max_{m=1,\cdots,M} |\sqrt{\sigma_1} {\mathbf v_m^H \mathbf{Q}(t)^H \mathbf{w}^\star }|^2 \nonumber\\
     &~~~~~~~~~~~~~~~~~~~~~~~+  \sum_{k=2}^{N_U} |\sqrt{\sigma_k} \mathbf v_m^H \mathbf{Q}(t)^H \mathbf{u}_k |^2, \label{appendix_pmi}
\end{align}
where $\{\mathbf{w}^\star, \{\mathbf u_k\}_{k=2}^{N_U}\}, \{\sigma_k \}_{k=1}^{N_U}$ ($\sigma_1 \geq \sigma_2 \geq \cdots \geq \sigma_{N_U}$) are the eigenvectors and eigenvalues of $\mathbf{C}$, respectively. From \eqref{appendix_pmi}, it is obvious that the PMI $m_0(t)$ is supposed to be chosen from the range of $[1, M]$ to align $\mathbf{v}_{m_0(t)}$ with $\mathbf{Q}(t)^H \mathbf{w}^\star $ for approaching the upper bound value of $\sigma_1$, i.e.,
\begin{align}
 &|{\mathbf v_{ m_0(t)}^H \mathbf{Q}(t)^H \mathbf{w}^\star }| \rightarrow 1;\\
 &|{\mathbf v_{ m_0(t)}^H \mathbf{Q}(t)^H \mathbf{u}_k}| \rightarrow 0, ~~ k =2,\cdots, N_U. 
\end{align}
However, due to the limited quantization quality of Type I codebook, it is often challenging for $\mathbf{v}_{m_0(t)}$ to precisely align with $\mathbf{Q}(t)^H \mathbf{w}^\star $, resulting in  $|{\mathbf v_{ m_0(t)}^H \mathbf{Q}(t)^H \mathbf{w}^\star }| \neq 1$ and the difficulty of bounding $|{\mathbf v_{ m_0(t)}^H \mathbf{Q}(t)^H \mathbf{u}_k}|$.

To tackle this challenge, we consider the decomposition as follows:
\begin{align}
    \mathbf{v}_{m_0(t)} =a_t\mathbf{Q}(t)^H\mathbf{w}^\star + \left(\sqrt{1-a_t^2}\right)\mathbf{o}, \label{v_decompose}
\end{align}
where $a_t =  |{\mathbf v_{ m_0(t)}^H \mathbf{Q}(t)^H \mathbf{w}^\star }|$ and unit-norm vector $\mathbf{o}$ is orthogonal to $\mathbf{Q}(t)^H\mathbf{w}^\star$. By introducing this decomposition into the term $|{\mathbf v_{ m_0(t)}^H \mathbf{Q}(t)^H \mathbf{u}_k}|$, we can yield
\begin{align}
    |{\mathbf v_{ m_0(t)}^H \mathbf{Q}(t)^H \mathbf{u}_k}|^2 = \left( {1-a_t^2}\right)  | {\mathbf{o}^H}&\mathbf{Q}(t)^H\mathbf{u}_k |^2,\nonumber\\
    &\quad \forall k = 2,\cdots,N_U.\label{v_decompose_2}
\end{align} 
In particular, equation \eqref{v_decompose_2} shows that the bound of $|{\mathbf v_{ m_0(t)}^H \mathbf{Q}(t)^H \mathbf{u}_k}|^2$ is bounded by alignment coefficient $a_t^2$ and bias magnitude $\left| {\mathbf{o}^H}\mathbf{Q}(t)^H\mathbf{u}_k\right|^2$. Though the precise values of $a_t^2$ and $\left| {\mathbf{o}^H}\mathbf{Q}(t)^H\mathbf{u}_k\right|^2$ is inaccessible, their value distribution can be obtained under the assumption that $\{\mathbf{w}^\star, \{\mathbf u_k\}_{k=2}^{N_U}\}$ are orthogonal random variables on sphere $\mathbb{S}^{N_A}$.

\textcolor{black}{Notice that $a^2_t$ can be equivalently rewritten as $\max(|\mathbf{v}_1^H\mathbf{Q}(t)^H\mathbf{w}^\star|^2,\cdots,|\mathbf{v}_M^H\mathbf{Q}(t)^H\mathbf{w}^\star|^2)$, where all terms $\{|\mathbf{v}_m^H\mathbf{Q}(t)^H\mathbf{w}^\star|^2\}_{m=1}^M$ are i.i.d variables.  Particularly, since $\mathbf{w}^\star$ is random variable on $\mathbb{S}^{N_A}$, $\mathbf{Q}(t)^H\mathbf{w}^\star$ is naturally uniform variable on $\mathbb{S^{N_P}}$. Furthermore, by changing the coordinate system, fixed vector $\mathbf{v}_m$ can be treated as $\mathbf{e}\triangleq(1,0,\cdots,0)\in\mathbb{C}^{N_P}$. Such a transform has no affect on statistic characteristic of $\mathbf{Q}(t)^H\mathbf{w}^\star$. According to \cite{fang2018symmetric}, we can conclude that $|\mathbf{v}_m^H\mathbf{Q}(t)^H\mathbf{w}^\star|^2$ follows the beta distribution $\mathrm{Beta}(\frac{1}{2},\frac{N_P}{2})$. By leveraging the order statistic theory, the cumulative distribution function of $a_t^2$ is $[I(\frac{1}{2},\frac{N_P}{2})]^M$, where $I$ is regularized incomplete beta function. }

In addition, notice that  $\mathbf{w}^\star$ are orthogonal to $\mathbf{Q}(t){\mathbf{o}}$ and any $\mathbf{u}_k$. Therefore, by leveraging the spherical symmetry property, we can transform the coordinate system so that $\mathbf{Q}(t){\mathbf{o}}$ corresponds to a random vector $\mathbf{x}$ on the sphere $\mathbb{S}^{N_A-1}$ and $\mathbf{u}_k$ corresponds to $\mathbf{e}_1 \triangleq(1,0,\cdots,0)\in\mathbb{C}^{N_A-1}$. Similalry, according to \cite{fang2018symmetric}, $|\mathbf{e}_1^H\mathbf{x}|^2$, sharing the same distribution of  $\left| {\mathbf{o}^H}\mathbf{Q}(t)^H\mathbf{u}_k\right|^2$, follows a beta distribution $\mathrm{Beta}(\frac{1}{2},\frac{N_A-1}{2})$, with mean $\mu_{(\text{bm})} = \frac{1}{N_A}$ and variance $\sigma^2_{(\text{bm})} = \frac{2(N_A-1)}{N_A^3 + 2N_A}$.

Therefore, based on Chebyshev's theorem, there is 
\begin{align}
    &\mathbb{P}\left(\left|a^2_t- \mu_{(\text{ac})}\right|<\epsilon_{(\text{ac})}\right) = \mathbb{P}\left(\left|(1-a^2_t)-(1 -\mu_{(\text{ac})})\right|<\epsilon_{(\text{ac})}\right)\nonumber\\
    &~~~~~~~~~~~~~~~~~~~~~~~~~~~~~~~\geq 1-\frac{\sigma_{(\text{ac})}^2}{\epsilon^2_{(\text{ac})}},\label{dis_1}\\ 
    &\mathbb{P}\left(\left|\left| {\mathbf{o}^H}\mathbf{Q}(t)^H\mathbf{u}_k\right|^2 - \mu_{(\text{bm})}\right|<\epsilon_{(\text{bm})}\right)\geq 1-\frac{\sigma^2_{(\text{bm})}}{\epsilon^2_{(\text{bm})}}, \label{dis_2}
\end{align}
where $\epsilon_{(\text{ac})}$ and $\epsilon_{(\text{bm})}$ are positive. By multiplying \eqref{dis_1} and \eqref{dis_2}, we can conclude that 
\begin{align}
    &\mathbb{P}\left( \left||{\mathbf v_{ m_0(t)}^H \mathbf{Q}(t)^H \mathbf{u}_k}|^2 - (1-\mu_{(\text{ac})})\mu_{(\text{bm})}\right| <\epsilon \right) \nonumber\\
    &\geq \mathbb{P}\left(\left|(1-a^2_t)-(1 -\mu_{(\text{ac})})\right|<\epsilon_{(\text{ac})}\right) \times \nonumber\\
    &~~~~~~~~~~~\mathbb{P}\left(\left|\left| {\mathbf{o}^H}\mathbf{Q}(t)^H\mathbf{u}_k\right|^2 - \mu_{(\text{bm})}\right|<\epsilon_{(\text{bm})}\right)\nonumber\\
    &\geq \left(1-\frac{\sigma_{(\text{ac})}^2}{\epsilon^2_{(\text{ac})}}\right)\left(1-\frac{\sigma^2_{(\text{bm})}}{\epsilon^2_{(\text{bm})}}\right),
\end{align}
where $\epsilon = (1-\mu_{(\text{ac})})\epsilon_{(\text{bm})} + \mu_{(\text{bm})}\epsilon_{(\text{ac})} + \epsilon_{(\text{ac})}\epsilon_{(\text{bm})}$. Moreover, due to the independence between $\{|{\mathbf v_{ m_0(t)}^H \mathbf{Q}(t)^H \mathbf{u}_k}|^2\}_{k=2}^{N_U}$, we can conclude that 
\begin{align}
     &\mathbb{P}\left( |\delta_t^2 - \sum_{k=2}^{N_U}\sigma_k(1-\mu_{(\text{ac})})\mu_{(\text{bm})}| <\sum_{k=2}^{N_U}\sigma_k\epsilon \right) \geq \nonumber\\
     & ~~~~~~~~~~~~~~~~~~~~~~~~~~ \left((1-\frac{\sigma_{(\text{ac})}^2}{\epsilon^2_{(\text{ac})}})(1-\frac{\sigma^2_{(\text{bm})}}{\epsilon^2_{(\text{bm})}})\right)^{N_U-1}.
\end{align}

\subsection{The equivalence of problem \eqref{12} and \eqref{EQ:14}}
\label{proof:equivalence}
By expanding the square term, problem \eqref{12} can be rewritten as follows:
\begin{align}
    \min_{{\mathbf{{w}}}} \sum_{t=1}^{T} &\mathbf{{w}}^H \mathbf{Q}(t) \mathbf{v}_{m_0(t)}\mathbf{v}_{m_0(t)}^{H} \mathbf{Q}(t)^H {\mathbf{w}} \nonumber \\
    &- 2\sqrt{\eta(t)}  |\mathbf{v}_{m_0(t)}^{H} \mathbf{Q}(t)^H {\mathbf{w}}| + \lambda\|{\mathbf{w}}\|^2. \label{EQ:13}
\end{align}
Therefore, we need to prove the equivalence between \eqref{EQ:13} and \eqref{EQ:14}, which can be further reorganized as 
\begin{align}
    |\mathbf{v}_{m_0(t)}^{H} \mathbf{Q}(t)^H {\tilde{\mathbf{w}}}| =  \Re\{\mathbf{v}_{m_0(t)}^{H} \mathbf{Q}(t)^H {\tilde{\mathbf{w}}}\exp(j\tilde{\phi_t}^\star)\}, ~~~
    \forall \tilde{\mathbf{w}},
\end{align}
where
\begin{align}
	{\tilde{\phi_t}^\star}  &= \arg\min_{{\phi_t}}\tilde{\mathbf{w}}^H \mathbf{Q}(t) \mathbf{v}_{m_0(t)}\mathbf{v}_{m_0(t)}^{H} \mathbf{Q}(t)^H {\tilde{\mathbf{w}}} \nonumber \\
 &~~~~~~~~~~~~~~~~- 2\sqrt{\eta(t)}  \Re\{\mathbf{v}_{m_0(t)}^{H} \mathbf{Q}(t)^H {\tilde{\mathbf{w}}}\exp(j\phi_t)\} \nonumber\\
 &~~~~~~~~~~~~~~~~~~~~+ \lambda\|{\tilde{\mathbf{w}}}\|^2  \nonumber \\
 & = \arg\min_{{\phi_t}} - 2\sqrt{\eta(t)}  \Re\{\mathbf{v}_{m_0(t)}^{H} \mathbf{Q}(t)^H {\tilde{\mathbf{w}}}\exp(j\phi_t)\} \nonumber \\
 & = \arg\max_{{\phi_t}} \Re\{\underbrace{\mathbf{v}_{m_0(t)}^{H} \mathbf{Q}(t)^H {\tilde{\mathbf{w}}}}_{\triangleq \mathbf{a}}\exp(j\phi_t)\} \nonumber\\
 & = \arg\max_{{\phi_t}} \Re\{ (\Re\{\mathbf{a}\}+j \Im\{\mathbf{a}\}) \times (\cos(\phi_t) + j \sin(\phi_t))\} \nonumber\\
 & = \arg\max_{{\phi_t}}  \Re\{\mathbf{a}\}\cos(\phi_t)-\Im\{\mathbf{a}\} \sin(\phi_t).
 \label{App_III_1_2}
\end{align} 
Based on the equation \eqref{App_III_1_2}, it can conclude that 
\begin{align}
	\sin(\tilde{\phi_t}^\star) &= -\frac{\Im\{\mathbf{a}\}}{\sqrt{\Re\{\mathbf{a}\}^2+ \Im\{\mathbf{a}\}^2}} \nonumber \\
	\cos(\tilde{\phi_t}^\star) & = \frac{\Re\{\mathbf{a}\}}{\sqrt{\Re\{\mathbf{a}\}^2+ \Im\{\mathbf{a}\}^2}}.\label{App_III_1_3}
\end{align}
Therefore, for any given $\tilde{\mathbf{w}}$, it holds that 
\begin{align}
	&\Re\{ \mathbf{v}_{m_0(t)}^{H} \mathbf{Q}(t)^H {\tilde{\mathbf{w}} \exp(j\phi_t}^\star)\} \nonumber\\
 &= \sqrt{\Re\{\mathbf{v}_{m_0(t)}^{H} \mathbf{Q}(t)^H \tilde{\mathbf{w}}\}^2+\Im\{ \mathbf{v}_{m_0(t)}^{H} \mathbf{Q}(t)^H \tilde{\mathbf{w}}\}^2} \nonumber\\
	& =  |\mathbf{v}_{m_0(t)}^{H} \mathbf{Q}(t)^H {\tilde{\mathbf{w}}} |. \label{App_III_1_4}
\end{align}

Based \eqref{App_III_1_4}, the equivalence of problem \eqref{12} and \eqref{EQ:14} can be concluded.

\subsection{The proof of \textbf{Proposition \ref{pro:2}}}
\label{proof:proposition2}
Given $ {\phi_t} = \hat{\phi}_t$, optimization problem \eqref{EQ:14} can be rewritten as 
\begin{align}
    \min_{\mathbf{w} } & \sum_{t=1}^{T} \Bigg({\mathbf{w} }^H \mathbf{Q}(t) \mathbf{v}_{m_0(t)} (\overline{\exp(j\hat{\phi}_t)}) \nonumber\\
    &~~~~~~~~~~~~\times \exp(j\hat{\phi}_t) \mathbf{v}_{m_0(t)}^H \mathbf{Q}(t)^H \mathbf{w}  \Bigg)\nonumber \\ 
   & ~~~ - 2\Re{ \left\{ \left( \sum_{t=1}^T \sqrt{\eta(t)} (\overline{\exp(j\hat{\phi}_t )}) \mathbf{Q}(t) \mathbf{v}_{m_0(t)} \right)^H \mathbf{w}  \right\}} \nonumber\\
   &~~~~~~+ \lambda \| \mathbf{w}\|. \label{pro_2_eq2}
\end{align}
By introducing $\mathbf{y}=\left[\sqrt{\eta(1)}, \sqrt{\eta(2)}, \cdots, \sqrt{\eta(T)}\right]^{T}$ and $\mathbf{\Phi}$ defined by \eqref{likelihood_PHI}, problem \eqref{pro_2_eq2} can be reformulated as 
\begin{align}
    \min_{\mathbf{w}}    \mathbf{w}^H \mathbf{\Phi} \mathbf{\Phi}^H \mathbf{w} - 2 \Re{ \{ \mathbf{y}^H \mathbf{\Phi} \mathbf{w} \} } + \lambda \| \mathbf{w} \|^2,\label{pro_2_eq3}
\end{align}
which can be further rewritten as
\begin{align}
    \min_{\mathbf{w}} & \| \mathbf{y} - \mathbf{\Phi} \mathbf{w}\|^2 + \lambda \| \mathbf{w}\|^2.\label{pro_2_eq4}
\end{align}

Based on Bayes' theorem, the log posterior distribution is proportional to the product of likelihood \eqref{likelihood} and prior distribution \eqref{prior}, that is
\begin{align}
    \log p(\mathbf{w} \mid \mathbf{y}; \alpha,\beta) \propto \ -\frac{\beta}{2} \| \mathbf{y} - \mathbf{\Phi} \mathbf{w} \|^2 - \frac{\alpha}{2} \| \mathbf{w} \|^2 . \label{pro_2_eq5}
\end{align}
Therefore, the maximum a posterior estimation problem can be formulated as:
\begin{align}
    \max_{\mathbf{w}}  -\frac{\beta}{2}\| \mathbf{y} - \mathbf{\Phi} \mathbf{w} \|^2 - \frac{\alpha}{2} \| \mathbf{w}\|^2.\label{pro_2_eq6}
\end{align}
Furthermore, since the introduction of a scalar multiplier does not change the solution, we can reformulate problem \eqref{pro_2_eq6} as 
\begin{align}
    \min_{\mathbf{w}} \frac{2}{\beta} \left( \frac{\beta}{2}\| \mathbf{y} - \mathbf{\Phi} \mathbf{w} \|^2 + \frac{\alpha}{2} \| \mathbf{w} \|^2\right),
\end{align}
which is equivalent to \eqref{pro_2_eq4} if $\lambda = \frac{\alpha}{\beta}$.

\bibliography{reference} 
\bibliographystyle{IEEEtran}

\begin{IEEEbiography}[{\includegraphics[width=1in,height=1.25in, clip,keepaspectratio]{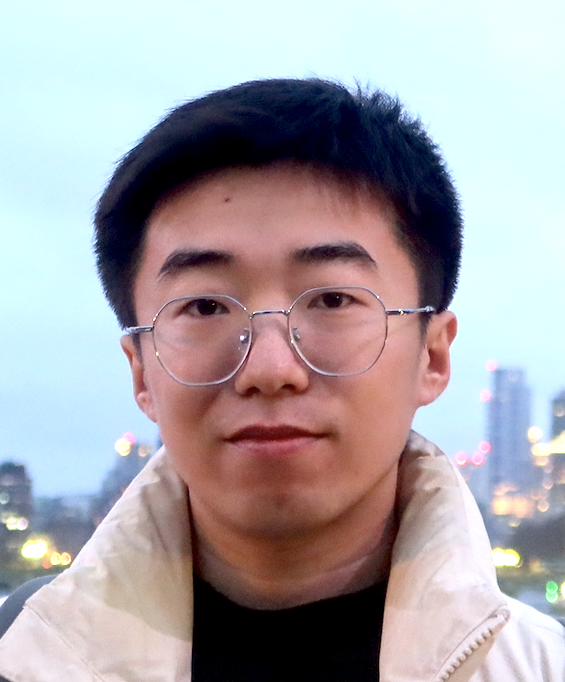}}]{Kai Li} received the B.S. degree in mathematics from Huazhong University of Science and Technology, Wuhan, China, in 2018. He is working toward the Ph.D. degree in computer information and engineering with the School of Science and Engineering at The Chinese University of Hong Kong, Shenzhen. His research interests include optimization, beamforming, and wireless communication.
\end{IEEEbiography}

\begin{IEEEbiography}[{\includegraphics[width=1in,height=1.25in, clip,keepaspectratio]{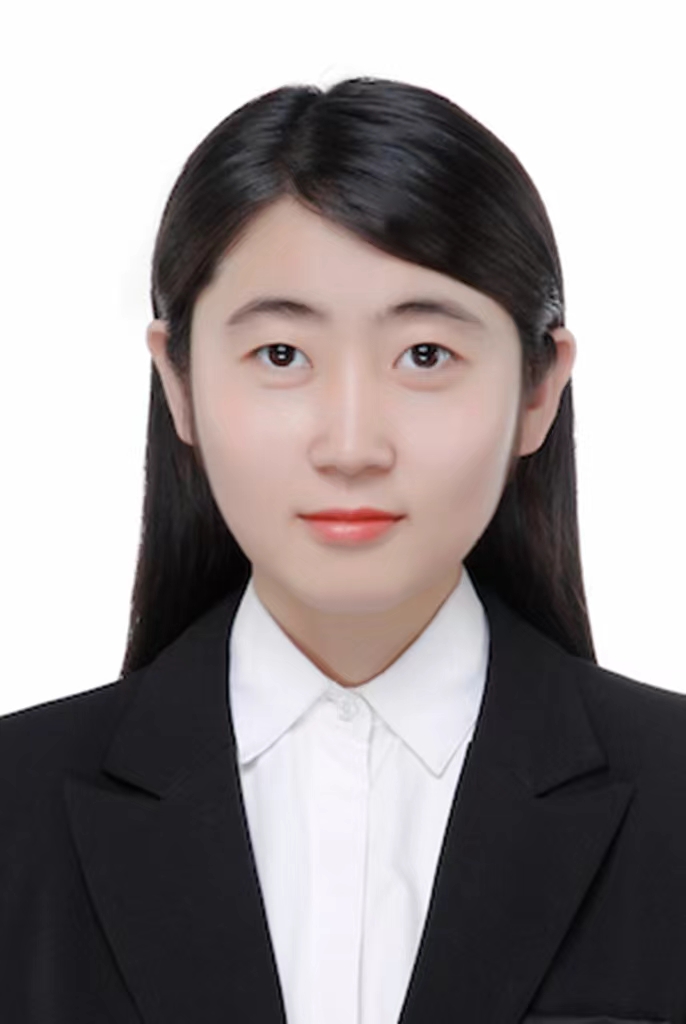}}]{Ying Li} received the B.Eng. degree from Shandong University in 2019 and M.Phil. degree from The Chinese University of Hong Kong, Shenzhen in 2022. She is currently pursuing a Ph.D. degree with the Department of Statistics and Actuarial Science at the University of Hong Kong, Hong Kong, SAR, China. Her research interests include Bayesian nonparametrics and variational auto-encoder.
\end{IEEEbiography}

\begin{IEEEbiography}[{\includegraphics[width=1in,height=1.25in, clip,keepaspectratio]{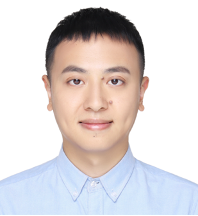}}]{Lei Cheng} (Member, IEEE) is currently ZJU Young Professor with the College of Information Science and Electronic Engineering, Zhejiang University, Hangzhou, China. He received the B.Eng. degree from Zhejiang University in 2013, and the Ph.D. degree from The University of Hong Kong in 2018. He was a Research Scientist in Shenzhen Research Institute of Big Data, The Chinese University of Hong Kong, Shenzhen, from 2018 to 2021. He is the author of the book ``Bayesian Tensor Decomposition for Signal Processing and Machine Learning'', Springer, 2023. He was a Tutorial Speaker in IEEE ICASSP 2023, Invited Speaker in ASA 2024 and IEEE COA 2014. He is now Associate Editor for \textit{Elsevier Signal Processing} and Young Editor for \textit{Acta Acustica}. His research interests are in Bayesian machine learning for tensor data analytics and interpretable machine learning for information systems.
\end{IEEEbiography}

\begin{IEEEbiography}[{\includegraphics[width=1in,height=1.25in, clip,keepaspectratio]{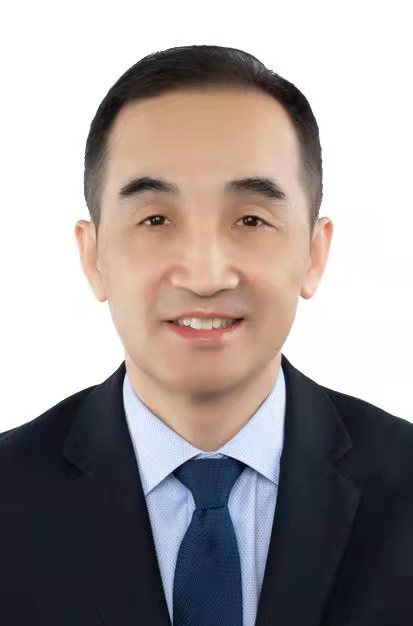}}]{Zhi-Quan Luo}(Fellow, IEEE) received the B.S. degree in mathematics from Peking University, Beijing, China, in 1984, and the Ph.D. degree in operations research from the Massachusetts Institute of Technology, Cambridge, MA, USA, in 1989. He is currently the Vice President (Academic) of The Chinese University of Hong Kong, Shenzhen, China, where he has been a Professor since 2014. He is currently the Director of the Shenzhen Research Institute of Big Data. His research interests include optimization, Big Data, signal processing and digital communication, ranging from theory, algorithms to design, and implementation. Professor Luo was the Chair of the IEEE Signal Processing Society Technical Committee on Signal Processing for Communications and Editor in Chief for IEEE transactions on signal processing during 2012–2014. He was an Associate Editor for many internationally recognized journals.

He is a Fellow of the Society for Industrial and Applied Mathematics. He was the recipient of the 2010 Farkas Prize from the INFORMS Optimization Society and 2018 Paul Y. Tseng Memorial Lectureship from the Mathematical Optimization Society. He was also the recipient of the three Best Paper Awards in 2004, 2009, and 2011, Best Magazine Paper Award in 2015, all from the IEEE Signal Processing Society, and 2011 Best Paper Award from the EURASIP. In 2014, he was elected to the Royal Society of Canada. He was elected to the Chinese Academy of Engineering as a Foreign Member in 2021, and was awarded the Wang Xuan Applied Mathematics Prize in 2022 by the China Society of Industrial and Applied Mathematics.
\end{IEEEbiography}

\end{document}